%% file: cma.tex
\documentclass[iop,revtex4]{emulateapj}
\slugcomment{Accepted by ApJ, June 3, 2016}
\usepackage{times,amsmath,url}

\shorttitle{YSOs in Canis Major}
\shortauthors{Fischer et al.}

\begin{document}

\title{A WISE Census of Young Stellar Objects in Canis Major}
\author{
William J. Fischer$^{1,3}$,
Deborah L. Padgett$^1$,
Karl L. Stapelfeldt$^2$,
and Marta Sewi{\l}o$^{1,3}$
}
\affil{
$^1$NASA Goddard Space Flight Center, Greenbelt, MD, USA; william.j.fischer@nasa.gov \\
$^2$Jet Propulsion Laboratory, California Institute of Technology, Pasadena, CA, USA
}
\footnotetext[3]{NASA Postdoctoral Program Fellow}

\begin{abstract}
With the Wide-Field Infrared Survey Explorer (WISE), we searched for young stellar objects (YSOs) in a 100~deg$^2$ region centered on the lightly studied Canis Major star forming region.  Applying stringent magnitude cuts to exclude the majority of extragalactic contaminants, we find 144 Class I candidates and 335 Class II candidates.  The sensitivity to Class II candidates is limited by their faintness at the distance to Canis Major (assumed as 1000 pc).  More than half the candidates (53\%) are found in 16 groups of more than four members, including four groups with more than 25 members each.  The ratio of Class II to Class I objects, $N_{\rm II}/N_{\rm I}$, varies from 0.4 to 8.3 in just the largest four groups.  We compare our results to those obtainable with combined Two Micron All Sky Survey (2MASS) and post-cryogenic Spitzer Space Telescope data; the latter approach recovers missing Class II sources.  Via a comparison to protostars characterized with the Herschel Space Observatory, we propose new WISE color criteria for flat-spectrum and Class 0 protostars, finding 80 and seven of these, respectively.  The distribution of YSOs in CMa OB1 is consistent with supernova-induced star formation, although the diverse $N_{\rm II}/N_{\rm I}$ ratios are unexpected if this parameter traces age and the YSOs are due to the same supernova.  Less massive clouds feature larger $N_{\rm II}/N_{\rm I}$ ratios, suggesting that initial conditions play a role in determining this quantity.
\end{abstract}


\section{INTRODUCTION}

\setcounter{footnote}{3}

Using the Spitzer Space Telescope, investigators catalogued thousands of young stellar objects (YSOs) in the nearest kiloparsec \citep[e.g.,][]{eva09,gut09,chu09,reb10,meg12,dun13}. Even so, the limited time available during the cryogenic mission led investigators to focus on especially massive or nearby molecular clouds.  With its all-sky survey at 3.4, 4.6, 12, and 22 \micron, the Wide-field Infrared Survey Explorer (WISE; \citealt{wri10}) can be used to identify young stellar objects with criteria similar to those established for Spitzer \citep{koe12,koe14} but over the entire sky.  This provides the opportunity to characterize instances of star formation outside the regions covered by targeted modern infrared surveys.  Newly identified YSOs may refine the initial stellar mass function, allow a better characterization of star and planet formation in regions with low initial gas densities, and identify nearby targets for high-resolution follow-up imaging and spectroscopy.

As a pilot study for more expansive searches, we present results here for a 100 deg$^2$ field centered on the Canis Major star forming region.  Star formation in the vicinity of Canis Major is centered amid the CMa OB1 and CMa R1 associations, 1--2$^\circ$ below the Galactic plane near a longitude of $224^\circ$.  \citet{gre08} provides a comprehensive review of pre-WISE studies of the region.  Optical images show extensive nebulosity dominated by the large \ion{H}{2} region S296, which extends 200\arcmin\ north to south \citep{sha59}, as well as S292, S297, and bright-rimmed clouds (BRC) 26 through 29 \citep{sug91}.  The unusual young binary star Z~CMa, in which one component is an FU Orionis object and the other is a Herbig star \citep{van04,hin13}, lies in the southern half of S296.  Another FU Ori star, V900 Mon \citep{rei12}, lies in the northwest of our search region in Lynds 1656. 

\citet{her77} argued that star formation in CMa R1 was induced by a supernova $\sim$ $5\times10^5$ years ago.  They reported a nearly complete ring of optical emission nebulosity defined by S296 and fainter regions to the south and east with a radius of $\sim$ 1.5$^\circ$.  In further support of this conclusion, the center of the ring is devoid of bright stars, and a runaway star HD 54662, presumably ejected by the supernova, lies on the northern edge of the ring.  The region is home to $5\times10^4$ $M_\sun$ of material distributed across 22 clouds as traced by $^{13}$CO gas \citep{kim04}.  The cloud surface densities range from 11 to 45 $M_\sun$ pc$^{-2}$ with a median of 21 $M_\sun$ pc$^{-2}$.

CMa OB1 is estimated to be at a distance of $990\pm50$ pc based on $uvby\beta$ photometry of its members \citep{kal00}.  \citet{gre08} recommends a distance of 1000 pc, which we adopt here.  There is no comprehensive study of star formation across the region, although \citet{reb13} published a catalog of Spitzer-identified YSO candidates in a rectangular region roughly 5\arcmin\ on a side and centered on BRC 27, and \citet{mal12} used near-infrared colors to find YSOs in a 7.5\arcmin\ square field around S297.  Parts of our search region, described below, have recently been mapped by the outer Galactic plane surveys of Spitzer and the Herschel Space Observatory.  \citet{eli13} analyzed the Herschel extended emission and point source population across the region.

Canis Major is intermediate in distance between two larger star-forming regions, Orion (420 pc; \citealt{meg12}) and Carina (2.3 kpc; \citealt{smi08}).  Star formation in Orion may be the result of 10 to 20 supernovae over a 12 Myr period that created the Orion / Eridanus superbubble \citep{bal08}.  Within the Orion A and B molecular clouds lie 3481 YSOs with infrared excesses \citep{meg16}.  The filament densities traced by $^{13}$CO are typically larger than those in Canis Major, ranging from 27 to 400 $M_\sun$ pc$^{-2}$ with a median of 85 $M_\sun$ pc$^{-2}$ \citep{nag98}.  In Carina, at 2.3 kpc, supernovae have not yet disrupted the environment, so star formation is regulated by stellar winds and ultraviolet radiation \citep{smi08}.  The region contains $>2\times10^4$ YSOs when the Spitzer infrared-detected population, limited to intermediate-mass YSOs at this distance, is extrapolated over the stellar initial mass function \citep{pov11}.

The YSOs we seek can be divided into classes based on the slopes of their infrared spectral energy distributions (SEDs).  In the widely used scheme put forth by \citet{lad87}, Classes I, II, and III are defined by their slopes $\alpha=\left(d\log\lambda S_\lambda\right)/\left(d\log\lambda\right)$, where $\lambda$ is the wavelength, $S_\lambda$ is the flux density at $\lambda$, and the slope is calculated between roughly 2 and 20 \micron.  The slope decreases with evolution from Class I to Class III due to the reduction in circumstellar material as a YSO approaches the main sequence. \citet{gre94} added a flat-spectrum class between Classes I and II, and \citet{and93} added a Class 0, redder than Class I, in which $>0.5\%$ of the source luminosity is emitted beyond 350 \micron.  Class 0 and I sources have $\alpha\ge0.3$, flat-spectrum sources have $-0.3\le\alpha<0.3$, Class II sources have $-1.6\le\alpha<-0.3$, and Class III sources have $\alpha<-1.6$ \citep{dun14}.  Another diagnostic is the bolometric temperature $T_{\rm bol}$, which is the effective temperature of a blackbody with the same flux-weighted mean frequency as the observed SED \citep{mye93}.  The temperatures 70 K, 650 K, and 2800 K are the widely adopted divisions between Classes 0 and I, I and II, and II and III \citep{che95}.

In the current study, we are limited to the WISE infrared bands (3.4 -- 22 \micron).  This inhibits our ability to distinguish among Class 0, I and flat-spectrum sources, so we limit ourselves initially to identifying potential Class I and II objects.  In addition, the relatively low sensitivity at 22 \micron\ prevents secure identifications of Class III sources at the distance of the Canis Major clouds.  Physically, Class I and II sources are expected to correspond to cases in which, respectively, a dusty protostellar envelope and a dusty circumstellar disk dominate the infrared emission.

In this paper, Section 2 describes the data, Section 3 describes our YSO selection technique and candidate list, and Section 4 contains an analysis of the YSO clustering properties. Section 5 shows how the results obtained with only AllWISE data products can be augmented with data from 2MASS, the Spitzer post-cryogenic mission, and Herschel; it includes criteria for identifying flat-spectrum and Class 0 YSOs in WISE.  Section 6 discusses the ratio of Class II to Class I YSOs in groups across CMa OB1 and compares the groups to those in other star-forming regions, and Section 7 offers our conclusions.

\section{DATA}

WISE is a 40 cm telescope in low-Earth orbit that surveyed the sky at 3.4, 4.6, 12, and 22 \micron\ (bands $W1$, $W2$, $W3$, and $W4$) with nominal angular resolutions in the respective bands of 6.1\arcsec, 6.4\arcsec, 6.5\arcsec, and 12.0\arcsec\ \citep{wri10}.  The mission launched on 2009 December 14 and initially performed an all-sky survey in all four bands. Depletion of its cryogen forced the end of $W4$ imaging in 2010 August and the end of $W3$ imaging in 2010 September.  A subsequent survey NEOWISE \citep{mai11} continues to image the sky in $W1$ and $W2$.  The AllWISE catalog combines data from the cryogenic and post-cryogenic phases of the initial WISE mission, covering the approximately one-year period from 2010 January to 2011 January.  It is searchable via IRSA,\footnote{http://irsa.ipac.caltech.edu/} the NASA/IPAC Infrared Science Archive, and contains over $7.4\times10^8$ objects.

\begin{figure*}
\includegraphics[width=\hsize,trim={0 0 0 15},clip]{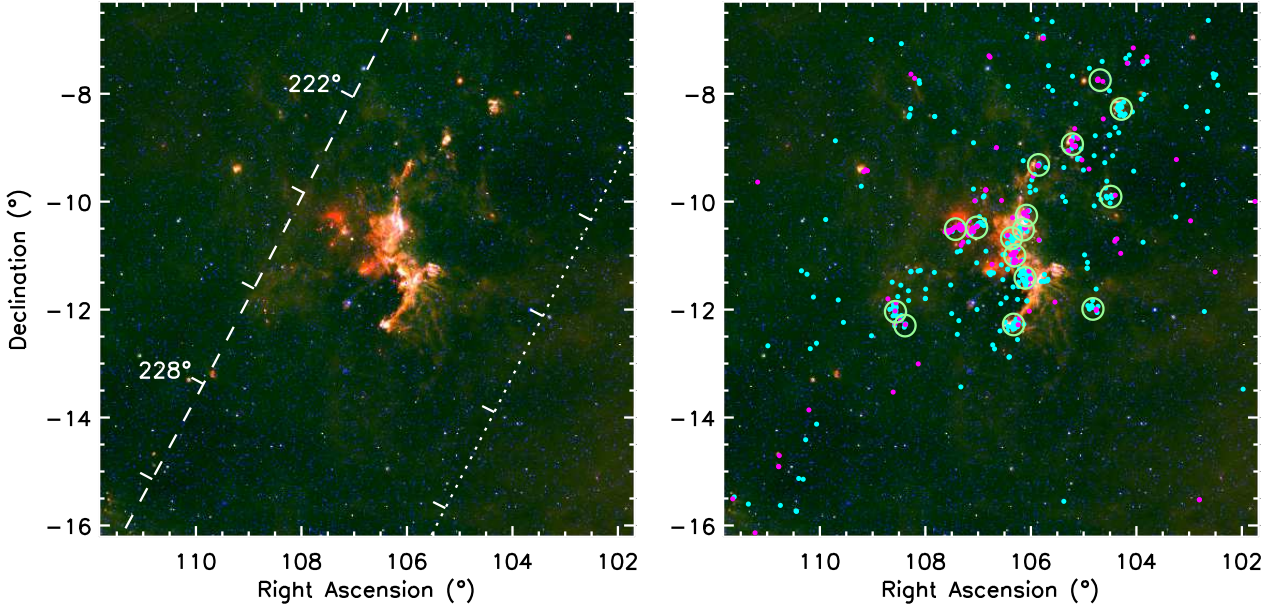}
\caption{{\em Left:} WISE map of the $10^\circ \times 10^\circ$ square centered at $106.67^\circ$ right ascension and $-11.29^\circ$ declination.  Blue is 3.4 \micron, green is 12 \micron, and red is 22 \micron.  The dashed line shows $b=0^\circ$, the dotted line shows $b=-5^\circ$, and ticks mark intervals in Galactic longitude of 2$^\circ$ as labeled. {\em Right:} The same map with magenta dots to indicate the positions of Class I candidates, cyan dots to indicate the positions of Class II candidates, and green circles to mark groups with more than 4 members.\label{f.region}}
\end{figure*}

We searched AllWISE in the $10^\circ \times 10^\circ$ square centered at $106.67^\circ$ right ascension and $-11.29^\circ$ declination.  These are the approximate coordinates of HD 53974 (FN CMa), a star that is roughly at the center of the CMa OB1 association.  We used the polygon method in IRSA, which searches within the region set off by great circles that connect the specified vertices ($101.67^\circ$ and $111.67^\circ$ in right ascension, and $-16.29^\circ$ and $-6.29^\circ$ in declination).  In this case, the great circles at the northern and southern limits of the field deviate from lines of constant declination by up to 0.4\%, and the area searched is 98.16 square degrees, but these details do not affect the list of YSO candidates.  The left panel of Figure~\ref{f.region} shows a three-color map of the region at 3.4, 12, and 22 \micron\ created with Montage\footnote{http://montage.ipac.caltech.edu/} from tiles in the WISE All-Sky Atlas.

In the cryogenic mission, WISE obtained four-band photometry in this region between 2010 March 29 and 2010 April 11.  Additional $W1$ and $W2$ photometry was obtained during the post-cryogenic mission, between 2010 October 7 and 2010 October 20.

\section{SELECTION OF YOUNG STELLAR OBJECTS}

\subsection{Identifying Point Sources of Interest}

The AllWISE catalog contains 2,264,279 sources in the $\sim$~100~deg$^2$ region.  To eliminate the least secure point sources without excluding likely YSOs, we make the following initial cuts:
\begin{itemize}
\item We require a detection rather than an upper limit in $W1$ and $W2$.  This is accomplished by requiring that the catalog parameters {\em w1sigmpro} and {\em w2sigmpro} not be null.  This excludes 274,749 sources.  A small number of these excluded sources (92) are not upper limits; they are sources that were nominally detected but for which no useful brightness estimate could be made.\footnote{http://wise2.ipac.caltech.edu/docs/release/allsky/expsup/index.html}
\item We ignore sources flagged as diffraction spikes, halos, optical ghosts, or persistence artifacts (flags $D$, $H$, $O$, or $P$) in any band.  This excludes 112,456 additional sources.
\item With the {\em xscprox} parameter, we exclude sources that are within 30\arcsec\ of a 2MASS extended source.  This is intended to avoid contamination from resolved galaxies and excludes 1224 additional sources.  The 2MASS extended source catalog contains 2011 sources in the field; if the exclusion zones are non-overlapping, they cover 0.44 square degrees, or 0.45\% of the field.
\item To exclude any severely saturated sources, we require $W1>5$ mag.  This excludes 250 additional sources.
\end{itemize}

These steps leave 1,875,600 sources, or 83\% of the original count.  In their treatment, \citet{koe14} also considered the signal-to-noise ratio and the reduced $\chi^2$ of the profile-fit photometry measurement in each band, rejecting sources with $\chi^2$ larger than a band-dependent threshold that increases with the signal-to-noise ratio.  We experimented with such cuts but found that any such cut retains problematic sources while removing good ones.  We thus avoid such cuts at the risk of introducing some contamination.

We do not explicitly require detections in $W3$ and $W4$, although all of the YSO candidates are detected in one or the other.  Of the 479 candidates in the final list, 468 are detected in both of these bands, eight are detected only in $W3$, and three are detected only in $W4$.

\subsection{Identifying Class I and Class II Candidates}

From the list generated with the selection criteria of Section 3.1, we select YSO candidates based on their $W1-W2$ and $W2-W3$ colors according to a slightly modified version of the \citet{koe14} criteria, which are based on the colors of known YSOs in Taurus, extragalactic sources, and Galactic contaminants.  The modification simplifies the criteria while excluding a small number of sources that would have otherwise been considered Class II.  Class I YSOs satisfy all four of the following constraints:
\begin{align}
W2-W3 &> 2.0, \\
W2-W3 &< 4.5, \\
W1-W2 &> 0.46\times(W2-W3)-0.9,~{\rm and} \\
W1-W2 &> -0.42\times(W2-W3)+2.2,
\end{align}
while Class II YSOs satisfy these constraints:
\begin{align}
W1-W2 &> 0.25, \\
W1-W2 &< 0.71\times(W2-W3)-0.07, \\
W1-W2 &> -1.5\times(W2-W3)+2.1, \\
W1-W2 &> 0.46\times(W2-W3)-0.9,~{\rm and} \\
W1-W2 &< -0.42\times(W2-W3)+2.2.
\end{align}

\begin{figure}
\includegraphics[width=\hsize]{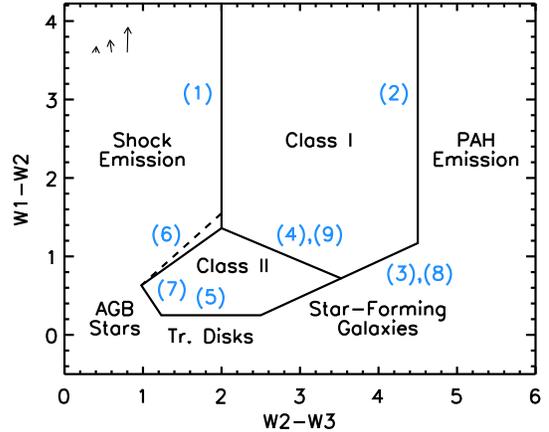}
\caption{WISE $W1-W2$ versus $W2-W3$ diagram showing the boundaries of the Class I and Class II color spaces.  Numbers indicate which of Equations 1 through 9 correspond to each line segment; the dashed line is the slighty different version of Equation 6 given in \citet[][their Equation 17]{koe14}.  The labels indicate where Class I and Class II YSOs are expected to lie as well as the major contributors to the population in other regions.  Arrows indicate reddening vectors for, from left to right, $A_K=0.4$, 0.8, and 2. \label{f.cc_annot}}
\end{figure}

Note that both classes satisfy the criterion described by Equations 3 and 8, while Class I and II YSOs fall on opposite sides of the line described by Equations 4 and 9.  These boundaries are illustrated in Figure~\ref{f.cc_annot}, where the numbers in parentheses show which line segment corresponds to each equation above.

We do not attempt to correct the colors for reddening due to dust along the line of sight, since, at WISE wavelengths, the reddening vectors are small, and their directions depend on the extinction \citep{mcc09}.  In Figure~\ref{f.cc_annot}, we show reddening vectors for $A_K=0.4$, 0.8, and 2.  These are derived from the WISE extinction law published by \citet{koe14}, which is based on interpolation of the \citet{mcc09} analysis of Spitzer/IRS spectra.  The vectors are not parallel due to the dependence of the extinction law on $A_K$.

Labels in Figure~\ref{f.cc_annot} indicate where Class I and Class II YSOs are expected to lie as well as which type of object is expected to be the main occupant of other regions of the color-color space.  ``Shock emission'' refers to resolved knots of shock emission that are prominent in $W2$ \citep{gut09,koe12}, while ``PAH emission'' refers to regions of polycyclic aromatic hydrocarbon emission with angular scales similar to the FWHM of the $W3$ point-spread function \citep{koe12}.  Objects with colors slightly redder than zero are predominantly asymptotic giant branch (AGB) stars, but other stars with weak infrared excesses, such as classical Be stars, lie there as well \citep{koe14}.  The transition disks lie at approximately $W1-W2=0$ mag but over a range of $W2-W3$ colors, and star-forming galaxies, including active galactic nuclei, lie at large $W2-W3$ and small $W1-W2$ due to their PAH content \citep{koe12}.

There are 6662 sources with the requisite Class I colors and 13,191 sources with the requisite Class II colors.  With the \citet{koe14} version of Equation 6, there would have been an additional 11 objects with Class II colors, and only two of these would satisfy the magnitude requirements discussed below.

\subsection{Filtering Likely Contaminants\label{s.filter}}

Most of the sources selected by the above color criteria are faint and uniformly distributed across the region.  YSOs, however, should be clustered near the sites of their formation. To deselect this distributed population of likely extragalactic sources with YSO colors, we further make $W1$ and $W4$ magnitude cuts:\  $W1<12$ mag or $W4<5$ mag.  Bluer YSOs satisfy the $W1$ cut, while the reddest YSOs fail the $W1$ cut but satisfy the $W4$ cut.

\begin{figure}
\includegraphics[width=\hsize]{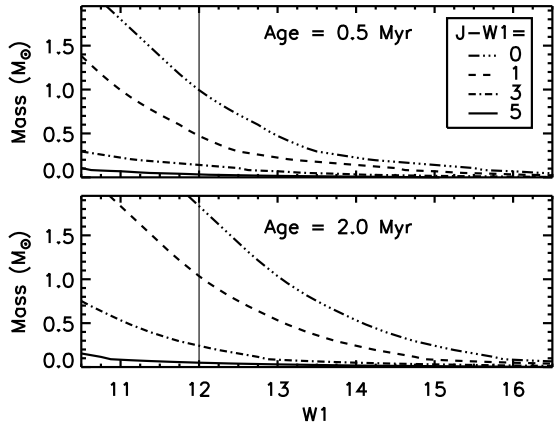}
\caption{YSO mass versus $W1$ magnitude.  From the \citet{bar15} models, we convert absolute $J$ magnitudes as a function of mass to apparent $J$ magnitudes, assuming a distance of 1000 pc.  The $J$ magnitude is then converted to a W1 magnitude for four colors:\ zero for photospheres and three additional values typical of the range for Class I and II YSOs.  The vertical lines show our cutoff of $W1=12$ mag and the implied mass limits as a function of color (line type) and age (top and bottom panels).\label{f.sens}}
\end{figure}

To establish the $W1$ cutoff magnitude, we considered the distribution in $W1$ of presumably extragalactic sources near the North Galactic Pole plotted by \citet{koe14}.  This falls off sharply as $W1$ goes from 13 mag to 12 mag; we chose 12 mag in the interest of keeping contamination to a minimum. For the $W4$ cut, we calculated the differential counts per unit 22 \micron\ flux density per unit area of sources with Class I colors, compared this to the same function for 24 \micron\ extragalactic sources in the Spitzer Wide-field Infrared Extragalactic (SWIRE) fields \citep{shu08}, and found that the Class I counts exceed the extragalactic counts above 79 mJy ($W4=5.0$ mag). 

The $W1$ cut comes at a cost in sensitivity to the Class II population, as shown in Figure~\ref{f.sens}.  To estimate the expected $W1$ magnitudes of YSOs, we take predicted $J$ magnitudes as a function of stellar mass and age from \citet{bar15}, scale them to 1000 pc, and explore the range of $J-W1$ colors found for Taurus YSOs \citep{reb11}.  For a cutoff of $W1<12$ mag at an age of 0.5 Myr, we are sensitive only to stellar masses $>1.0$ $M_\sun$ if $J-W1=0$ mag, $>0.47$ $M_\sun$ if $J-W1=1$ mag, and $>0.14$ $M_\sun$ if $J-W1=3$ mag.  For an age of 2 Myr, the bounds for these colors change to 1.8, 1.0, and 0.24 $M_\sun$.  The mass limits are rather stringent for sources with small $J-W1$, e.g., Class II YSOs, and less important for sources with larger $J-W1$, e.g., Class I YSOs.

In addition to the previously mentioned cuts, we also require $W1>6$ mag to remove red giants with infrared excesses.  Sources brighter than this limit tend to correspond to previously cataloged red giant stars.  Of the sources with the requisite colors, 146 Class I and 343 Class II sources satisfy the magnitude requirements.  If the magnitude requirements were relaxed to $W1<13$ mag or $W4<6$ mag, the counts would have nearly doubled to 244 Class I sources and 658 Class II sources.

We visually inspected these 489 sources in WISE and 2MASS images.  Candidates were classified with the following scheme:

\begin{itemize}
\item {\bf y}es; unambiguous detection; contamination likely insignificant; 292 sources;
\item {\bf b}inary; unambiguous detection; contamination from nearby point source(s) may be significant; 96 sources;
\item {\bf f}aint; unambiguous point source in $W1$ and $W2$; nothing visible in $W4$; no obvious contaminants in $W3$ or $W4$; 53 sources;
\item {\bf c}ontaminated; unambiguous point source in $W1$ and $W2$; may be a point source in $W3$ and $W4$, but probably contaminated by nearby point source or nebulosity; 38 sources;
\item {\bf n}o; point source not seen in any band; probably an artifact but not flagged; 10 sources.
\end{itemize}

The majority of the sources (60\%) have the {\bf y} code and are the most firm detections. Those with {\bf b}, {\bf f}, or {\bf c} codes (38\%) are likely YSOs, but the photometry may be less accurate.

\begin{figure}
\includegraphics[width=\hsize,trim={0 0 0 32.5},clip]{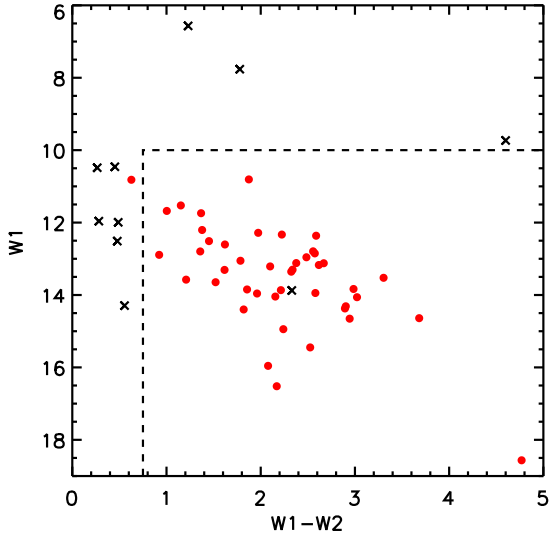}
\caption{WISE color-magnitude diagram showing $W1$ versus $W1-W2$ for the 54 sources that lack 2MASS counterparts and remain after point-source, color, and magnitude cuts.  Sources marked with $\times$ symbols have {\bf n} codes from visual inspection and are suspected unflagged artifacts.  Sources marked with filled circles have other inspection codes.  Those with $W1-W2<0.75$ mag or $W1<10$ mag are rejected on the basis that real sources with such WISE photometry should appear in 2MASS.\label{f.null2mass}}
\end{figure}

The two percent of sources that have code {\bf n} all lack 2MASS counterparts; that is, the AllWISE catalog parameter {\em tmass\_key} is null.  Sources without a 2MASS counterpart may be too faint at 2.2 \micron\ and shorter wavelengths to be detected by 2MASS, or they may be unflagged artifacts in the WISE images.  Of the 489 sources that remain at this step, Figure~\ref{f.null2mass} shows $W1$ versus $W1-W2$ for the 54 that lack 2MASS counterparts.  The ten sources with code {\bf n}, shown with $\times$ symbols, are almost all bright in $W1$ or near zero in $W1-W2$, suggesting they would have 2MASS $K$ magnitudes if they were real.  We thus eliminate from the catalog sources with a null {\em tmass\_key} and either $W1-W2<0.75$ mag or $W1<10$ mag.  This removes nine of the ten sources with code {\bf n} and removes one additional source with code {\bf b}.  We note that these criteria were motivated by inspection of sources that had already passed the other selection criteria; modifications may be needed to cull unflagged artifacts from more general source lists.

\begin{figure}
\includegraphics[width=\hsize]{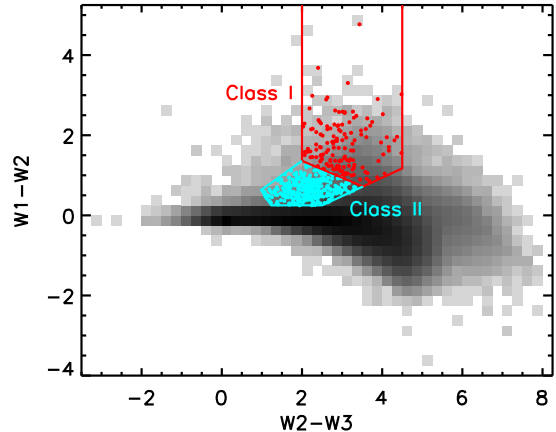}
\caption{WISE color-color diagram showing the locations of Class I (red) and Class II (cyan) YSO candidates that satisfy all criteria.  The shading indicates the distribution of objects that pass the point source criteria of Section 3.1; the lightest boxes contain only one object, and the darkest contains 81,380 objects.\label{f.cc}}
\end{figure}

Table~\ref{t.yso} lists the positions, 2MASS and WISE magnitudes and uncertainties, YSO classifications, and visual inspection codes of the 479 remaining candidates.  The right panel of Figure~\ref{f.region} shows their locations within the search field, and Figure~\ref{f.cc} compares their distribution on the $W1-W2$ versus $W2-W3$ diagram to that of all candidates that pass the point source criteria of Section 3.1.

Well-known YSOs do not necessarily appear in the final sample; for example, Z CMa is saturated and lacks reliable photometry, while the recent FU Ori outburst V900 Mon has Class I colors but is flagged as a halo artifact in all four bands.  Therefore, Table~\ref{t.yso} should be treated as a list of WISE-identified YSOs in the region, rather than a definitive list of every YSO in the region.

To estimate the level of contamination by red point sources that are not YSOs, we examined the region near Galactic longitude $\ell=228^\circ$, where the 12 \micron\ map made by the Infrared Astronomy Satellite (IRAS) shows little of the diffuse emission associated with star formation.  Since the contamination fraction may vary with Galactic latitude $b$, we examined the range $-6^\circ<b<3^\circ$, representative of the range covered by the 100 deg$^2$ search field.  We constructed a series of boxes perpendicular to the Galactic equator with centers ranging from $\ell=227.7^\circ$ to 229.4$^\circ$ in steps of 0.1$^\circ$ and counted the YSO candidates within them. These boxes have width $\Delta \ell=2^\circ$ and cover the range of Galactic latitude mentioned.

With many partially overlapping boxes, we reduce the influence of the precise location of a single box on the estimated contamination fraction.  The number of Class I candidates in each box ranges from 2 to 4, while the number of Class~II candidates ranges from 7 to 12.  Rescaled from 18 deg$^2$ to 100 deg$^2$, this implies that there are on average $16.0\pm5.4$ Class~I contaminants and $53.4\pm8.1$ Class II contaminants in the whole field; the uncertainties are the standard deviations in the counts from box to box. Repeating the exercise with $\Delta \ell$ of 1.5$^\circ$ or 2.5$^\circ$ yielded counts that were well within these uncertainties.  We thus estimate that $16.0/144=11\%$ of our Class I candidates and $53.4/335=16\%$ of our Class II candidates may be non-YSO point source contaminants.

\section{PROPERTIES OF STAR-FORMING GROUPS}

In this section we look at the spatial distribution of the YSO candidates.  There are a variety of methods to divide a population of point sources into clusters \citep{kuh14}, but there is no definitive best technique \citep{fei11}.  For our YSO candidates, we investigated the single, complete, and average linkage methods for agglomerative hierarchical clustering \citep{eve11}.  In these cases, YSOs are combined into progressively larger clusters, where, initially, the ``clusters'' are the individual YSOs.  At each step, the nearest two remaining clusters are combined.  After the YSOs have been combined into a single cluster, a critical length and a minimum membership number are chosen.  Links longer than the critical length are broken, and remaining clusters with at least the minimum number of members are retained.

The methods differ in how the distance between two clusters is defined.  For the single, complete, and average linkage methods, the intercluster distance is, respectively, the distance between the closest members of each cluster, the farthest members of each cluster, and the average distance between a member of one cluster and a member of the other.  Single linkage tends to chain together unrelated groups, complete linkage tends to find compact clusters of equal diameter, and average linkage is intermediate to the other cases \citep{eve11}.

For a given critical length, the three methods generate similar lists of the clustered versus distributed population.  They differ in that the average and complete methods tend to break clusters found by the single-linkage method into a few smaller clusters.  Further guidance as to which method best identifies physically related clusters of YSOs would require, for example, radial velocity data for the constituent YSOs to indicate which have a common origin.  For subsequent comparison with the \citet{gut09} catalog of clusters in the nearest kpc, we adopt their approach and use the single-linkage method.  We are primarily interested in a repeatable, quantitative division of the YSOs into clusters and a distributed population, similar to what one might pick out via visual inspection, and this approach satisfies that need.

\begin{figure}
\includegraphics[width=\hsize,trim={0 0 0 32.5},clip]{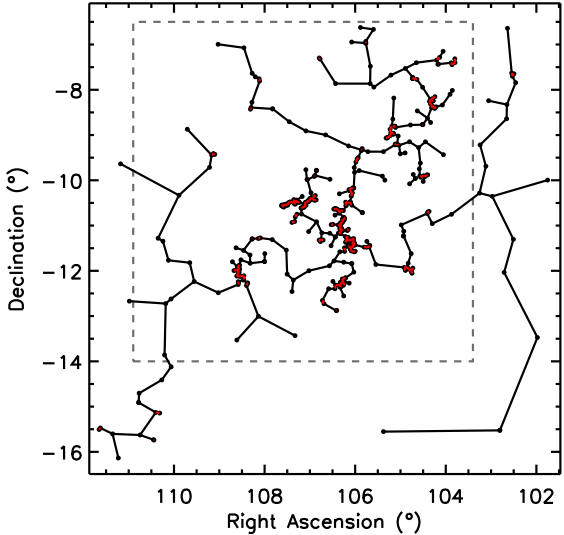}
\caption{The minimum spanning tree for the 479 YSO candidates, which are marked with circles. Branches shown in red are shorter than the 330\arcsec\ critical length and connect sources that belong to the same group.  The dashed gray box indicates the region shown in Figure~\ref{f.ratio}.\label{f.mst}}
\end{figure}

To implement the single-linkage method, first we calculate the minimum spanning tree (MST) for the population \citep{gut09,koe15}, shown in Figure~\ref{f.mst}.  The MST connects the candidates with the shortest total distance using no closed loops.  Techniques for determining the critical length at which to break the MST into clusters vary among authors.  \citet{sar15} discuss how the choice of critical length is partially dependent on the survey sensitivity and angular resolution as well as the true distribution of YSOs, and that different choices of critical length highlight different aspects of the clustering hierarchy in a region.  Our goal is to identify likely clusters over a large region rather than to present a detailed analysis of the substructure.  We find that the number of clusters with more than four members and their locations are unchanged for critical lengths ranging from 295\arcsec\ to 371\arcsec\ (1.43 to 1.80 pc).  We therefore choose the middle of this range, 330\arcsec\ (1.6 pc) as the critical length.  This approach yields 55 clusters of at least two members; however, we ignore the 39 with four or fewer members.  These choices avoid the unrealistic chains of objects that are a concern for the single-linkage method.

\citet{lad03} established a common definition of a cluster as a group of 35 or more physically related stars with a density exceeding 1.0 $M_\sun$ pc$^{-3}$.  One of the clusters we find has more than 35 candidates, and more likely do if we account for Class II objects not detected by WISE, but given this uncertainty, we henceforth use the term ``groups'' in all cases instead of ``clusters.''

\setcounter{table}{1}
\begin{deluxetable*}{lcrccccccccc}
\tablecaption{Properties of Groups with More Than Four Members\tablenotemark{1}\label{t.cluster}}
\tablewidth{\hsize}
\tablehead{ID & \multicolumn{2}{c}{Central Coordinates} & \multicolumn{2}{c}{YSO Counts} & \colhead{$N_{\rm II}/N_{\rm I}$} & \colhead{$A$ (pc$^2$)} & \multicolumn{2}{c}{$\sigma$ (pc$^{-2}$)} & \multicolumn{3}{c}{NN2$_{\rm med}$ (pc)} \\
\cline{2-3} \cline{4-5} \cline{8-9} \cline{10-12} \\ [-1.5ex]
\colhead{} & \colhead{RA (deg)} & \colhead{Dec (deg)} & \colhead{Class I} & \colhead{Class II} & \colhead{} & \colhead{} & \colhead{Class I} & \colhead{Class II} & \colhead{All} & \colhead{Class I} & \colhead{Class II}}
\startdata
00 & 107.43050 & $-$10.50928 &   29 &   12 &    0.4 &   11. &    2.7 &    1.1 &   0.13 &   0.20 &   0.86 \\
01 & 106.10981 & $-$11.41076 &    3 &   25 &    8.3 &   22. &    0.1 &    1.2 &   0.33 &   2.48 &   0.43 \\
02 & 107.02746 & $-$10.47745 &   11 &   17 &    1.6 &   13. &    0.8 &    1.3 &   0.27 &   0.08 &   0.34 \\
03 & 106.30586 & $-$10.99740 &   11 &   15 &    1.4 &   12. &    0.9 &    1.2 &   0.31 &   0.76 &   0.57 \\
04 & 108.56644 & $-$12.03924 &    3 &   15 &    5.0 &   19. &    0.2 &    0.8 &   0.28 &   2.64 &   0.59 \\
05 & 106.32906 & $-$12.27550 &    2 &   16 &    8.0 &   15. &    0.1 &    1.1 &   0.52 &   2.18 &   0.52 \\
06 & 105.21333 &  $-$8.93511 &    9 &    5 &    0.6 &   21. &    0.4 &    0.2 &   0.31 &   0.31 &   1.37 \\
07 & 106.36645 & $-$10.67523 &    1 &   13 &   13.  &   4.0 &    0.3 &    3.3 &   0.25 &  \nodata &   0.28 \\
08 & 104.28607 &  $-$8.28254 &    0 &   13 &   $\infty$ &   14. &    0 &    0.9 &   0.77 &  \nodata &   0.77 \\
09 & 106.08324 & $-$10.25760 &    5 &    6 &    1.2 &    6.2 &    0.8 &    1.0 &   0.18 &   0.29 &   0.18 \\
10 & 106.14224 & $-$10.51928 &    5 &    5 &    1.0 &    6.8 &    0.7 &    0.7 &   0.11 &   0.07 &   1.27 \\
11 & 104.82373 & $-$11.98775 &    2 &    7 &    3.5 &   11. &    0.2 &    0.7 &   0.71 &   0.37 &   0.81 \\
12 & 104.68740 &  $-$7.74870 &    5 &    1 &    0.2 &    1.4 &    3.6 &    0.7 &   0.16 &   0.19 &  \nodata \\
13 & 104.48487 &  $-$9.90643 &    1 &    5 &    5.0 &    8.4 &    0.1 &    0.6 &   0.58 &  \nodata &   0.62 \\
14 & 108.38586 & $-$12.29614 &    1 &    4 &    4.0 &    0.6 &    1.7 &    6.7 &   0.16 &  \nodata &   0.29 \\
15 & 105.85557 &  $-$9.31867 &    2 &    3 &    1.5 &    0.6 &    3.2 &    4.8 &   0.15 &   0.15 &   0.37
\enddata
\tablenotetext{1}{Assuming a distance of 1000 pc for $A$, $\sigma$, and NN2$_{\rm med}$.}
\end{deluxetable*}

Table~\ref{t.cluster} lists the properties of the 16 groups with more than four members.  We show the coordinates of the approximate center of each group, the number of Class I and Class II candidates $N_{\rm I}$ and $N_{\rm II}$ (where $N_{\rm YSO}=N_{\rm I}+N_{\rm II}$), and the ratio of the Class II to Class I counts.  Since the number of Class~II candidates is a lower limit (Section 3.3), this ratio is also a lower limit.  We next tabulate the area $A$ of each group, assuming a distance of 1000 pc.  This is the area of the group's convex hull (the smallest polygon containing all members that has all internal angles smaller than 180$^\circ$) adjusted upward for the fraction of members that are vertices.  The adjusted area is the hull area times $N_{\rm YSO}/N_{\rm in}$, where $N_{\rm in}$ is the number of members that are not vertices \citep{gut09,sch06}.

\setcounter{figure}{22}
\begin{figure}
\includegraphics[width=\hsize,trim={0 0 0 32.5},clip]{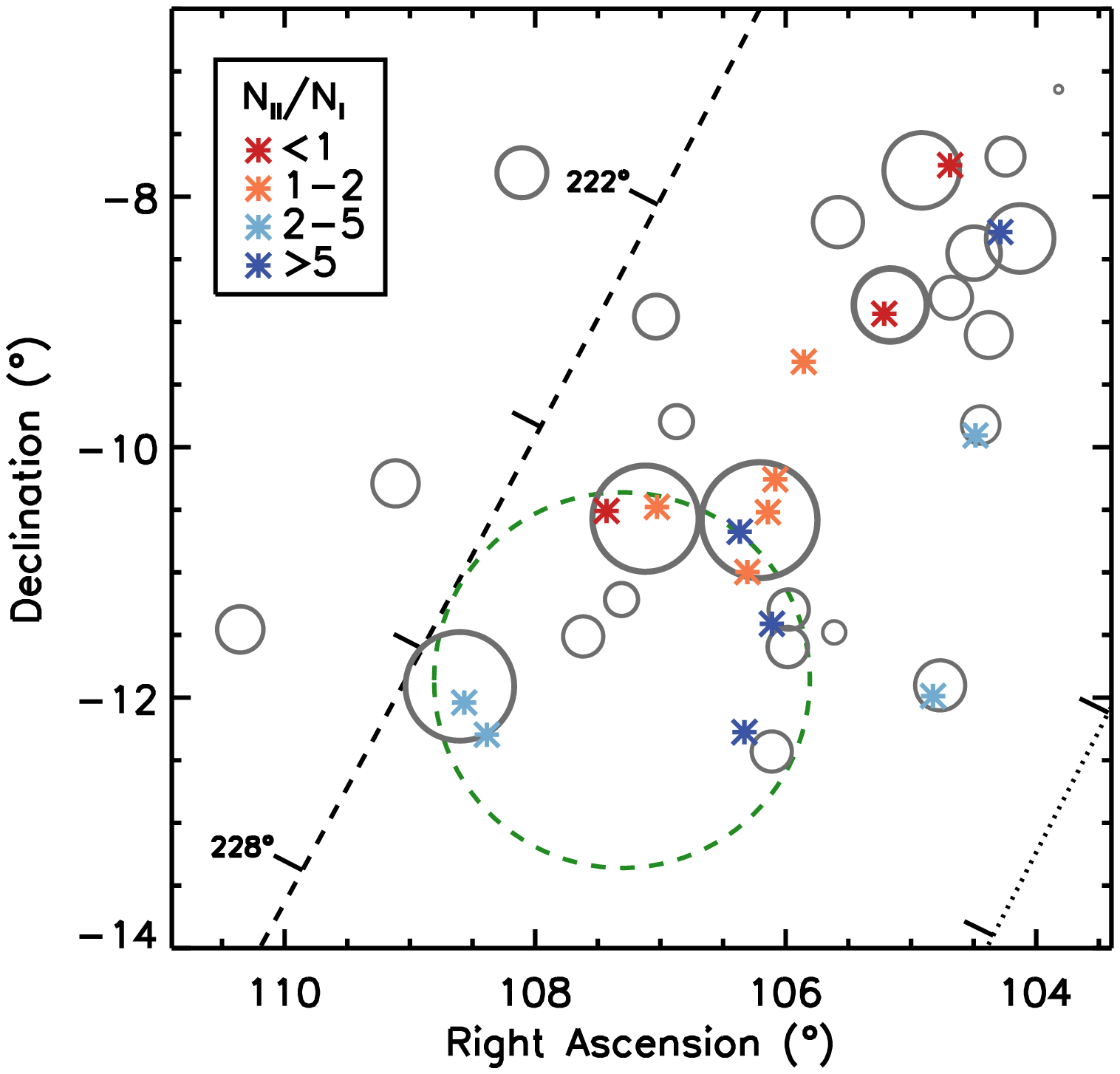}
\caption{Locations of groups (asterisks; this study) and $^{13}$CO clouds (circles; \citealt{kim04}).  Groups are colored based on their ratios of Class II sources to Class I sources.  Clouds are plotted as gray circles of the areas tabulated by \citeauthor{kim04}, and their line thicknesses are proportional to the logarithm of their enclosed masses.  The dashed green circle approximates the ring of optical and radio emission reported by \citet{her77}.  This is a $7.5^\circ \times 7.5^\circ$ cutout of the search field, as indicated in Figure~\ref{f.mst}.  To facilitate comparison with the \citeauthor{kim04}\ maps, a guide to Galactic coordinates is shown, as in Figure~\ref{f.region}.\label{f.ratio}}
\end{figure}

We then tabulate the surface density $\sigma$ of Class I and Class~II candidates and the median separation NN2$_{\rm med}$ of all identified YSOs and of the YSOs in each class.  The right panel of Figure~\ref{f.region} shows the locations of the 16 larger groups within the search field.  Figures~\ref{f.cluster0} through \ref{f.cluster15} show the location of each, a close-up view of each, and a schematic showing the local MST, the convex hull, and the locations of the Class I and Class II candidates.

The largest groups have widely varying ratios of Class II to Class I sources; the two largest have ratios of 0.4 and 8.3.  Figure~\ref{f.ratio} shows the locations of the groups with more than four members, coloring them according to their Class II to Class I ratios.  It also shows the locations and sizes of the $^{13}$CO clouds tabulated by \citet{kim04}.  The area of each circle is that listed in their Table 1, converted from pc$^2$ to deg$^2$ with the distances given.  We do not attempt to reproduce the shapes of the clouds.  The thickness of the line marking each circle is proportional to the logarithm of the tabulated cloud mass, where the masses range from 230 to 23000 $M_\sun$.

The YSO groups are generally found near the $^{13}$CO peaks.  The ten groups within half a degree of the supernova ring (dashed green circle in Figure~\ref{f.ratio}) correspond to CMa OB1 clouds tabulated in \citet{kim04}.  Elsewhere, Groups 08, 12, and 13 in the northwest correspond to clouds in G220.8$-$1.7.  Group 06, also in the northwest, corresponds to the cloud BFS 64, Group 11 in the southwest corresponds to a cloud in the southern filament of Orion, and Group 15 is not associated with any of their clouds.

The four most massive clouds, all more massive than 5000~$M_\sun$, host nine groups that contain 167 YSOs, 35\% of the total in the entire field, as well as additional YSOs that are not counted as members of a group.  Protostars tend to be associated with massive clouds, as the three most massive clouds, with masses exceeding $10^4$ $M_\sun$, host all four of the groups with more than five Class I YSOs.  The least massive clouds typically are not associated with groups of YSOs.  Of the 23 least massive clouds, with masses under 1500 $M_\sun$ (five of which are outside the boundaries of Fig.~\ref{f.ratio}), only four lie near YSO groups, and these YSO groups are dominated by Class II candidates.

There are 227 YSOs outside groups of more than four members.  This population features 54 Class I sources and 173 Class II sources, for a ratio $N_{\rm II}/N_{\rm I}=3.2$.  This is near the median of the group ratios, and it is larger than the ratio obtained when adding up all groups, $N_{\rm II}/N_{\rm I}=162/90=1.8$.

\citet{mye12} developed a method for converting the fraction of YSOs that are protostars to a group age.  While several assumptions are required to calculate absolute group ages, the groups can be {\em ordered} by likely age from their Class II to Class I ratios alone.  Making the same assumptions as \citet{mye12}, that there are as many undetected Class III YSOs as there are detected Class II YSOs and that the mean accretion time is 0.17 Myr, we find group ages ranging from 0.13 to 4.59 Myr, ignoring the group where there are no Class I detections and, thus, the method returns an infinite age.  The median group age is 1.35 Myr.  These ages increase if we account for our insensitivity to fainter Class II sources.  

In CMa OB1, the groups with the smallest ratios are roughly concentrated along the northern edge of the supernova ring, while those to the south have larger ratios.  The implied ages range from 0.23 to 4.59 Myr.  Interpretation of the II/I ratio as an age effect is at odds with a scenario where a single supernova is responsible for star formation in CMa OB1; we will explore this further in Section 6.1.  We compare the properties of these groups to those in the nearest kpc analyzed by \citet{gut09} in Section~\ref{s.comp}.

\section{AUGMENTING WISE YSO SEARCHES WITH 2MASS, SPITZER, AND HERSCHEL PHOTOMETRY}

In this section, we show how the results of our selection process, which uses only the AllWISE catalog, can be augmented with other catalogs of infrared photometry.  First, we compare the numbers of Class I and Class II YSOs recovered by our selection criteria to those recovered by criteria based on 2MASS and the two shortest-wavelength Spitzer channels.  Then, we use results from the Herschel Space Observatory to estimate the expected colors of flat-spectrum and Class 0 protostars in WISE.

\subsection{Comparison of WISE and Spitzer+2MASS Classifications in Group 00}

\citet{gut09} established a method of using photometry at $J$, $H$, $K$, 3.6 \micron, and 4.5 \micron\ to estimate the extinction toward YSOs and place them in Class I or II.  Here we use this method to identify and classify YSOs in Group 00, our most populous group.  This is particularly useful in estimating the number of Class II YSOs missed by our AllWISE-only search.

For the classification, we use photometry from 2MASS and the Galactic Legacy Infrared Mid-Plane Survey Extraordinaire 360 (GLIMPSE360) project \citep{whi11}, which surveyed the outer Galaxy at the 3.6 \micron\ and 4.5 \micron\ wavelengths available in Spitzer's post-cryogenic mission. GLIMPSE360 observed a strip across our search region, parallel to the Galactic equator with roughly $-2.3^\circ<b<0.5^\circ$, which contains Group 00.  We use the catalog (described as highly reliable) as opposed to the archive (described as more complete and less reliable), and we refer to the recovered objects as S2M candidates.  A thorough analysis of the Canis Major GLIMPSE360 observations appears in M.\ Sewi{\l}o et al.\ (in preparation), who present a method of selecting YSOs by combining GLIMPSE360 and WISE photometry.

We refer the reader to \citet{gut09} for the details of the classification process and present a brief overview here.  Objects with uncertainties less than 0.1 mag in the 2MASS $H$ and $K$ bands are considered.  The $J$ magnitude is also used if its uncertainty is $<0.1$ mag.  The foreground extinction for each object is determined by comparing its position in either $J-H$ versus $H-K$ space or $H-K$ versus $[3.6]-[4.5]$ space to a predetermined locus.  The colors are then dereddened accordingly, and objects are placed on a plot of $[K-[3.6]]_0$ versus $[[3.6]-[4.5]]_0$, where the subscript indicates the dereddened color.  Objects with moderately red colors are Class II, while those with more extreme colors are Class I.

\begin{figure}
\includegraphics[width=\hsize,trim={0 0 0 32.5},clip]{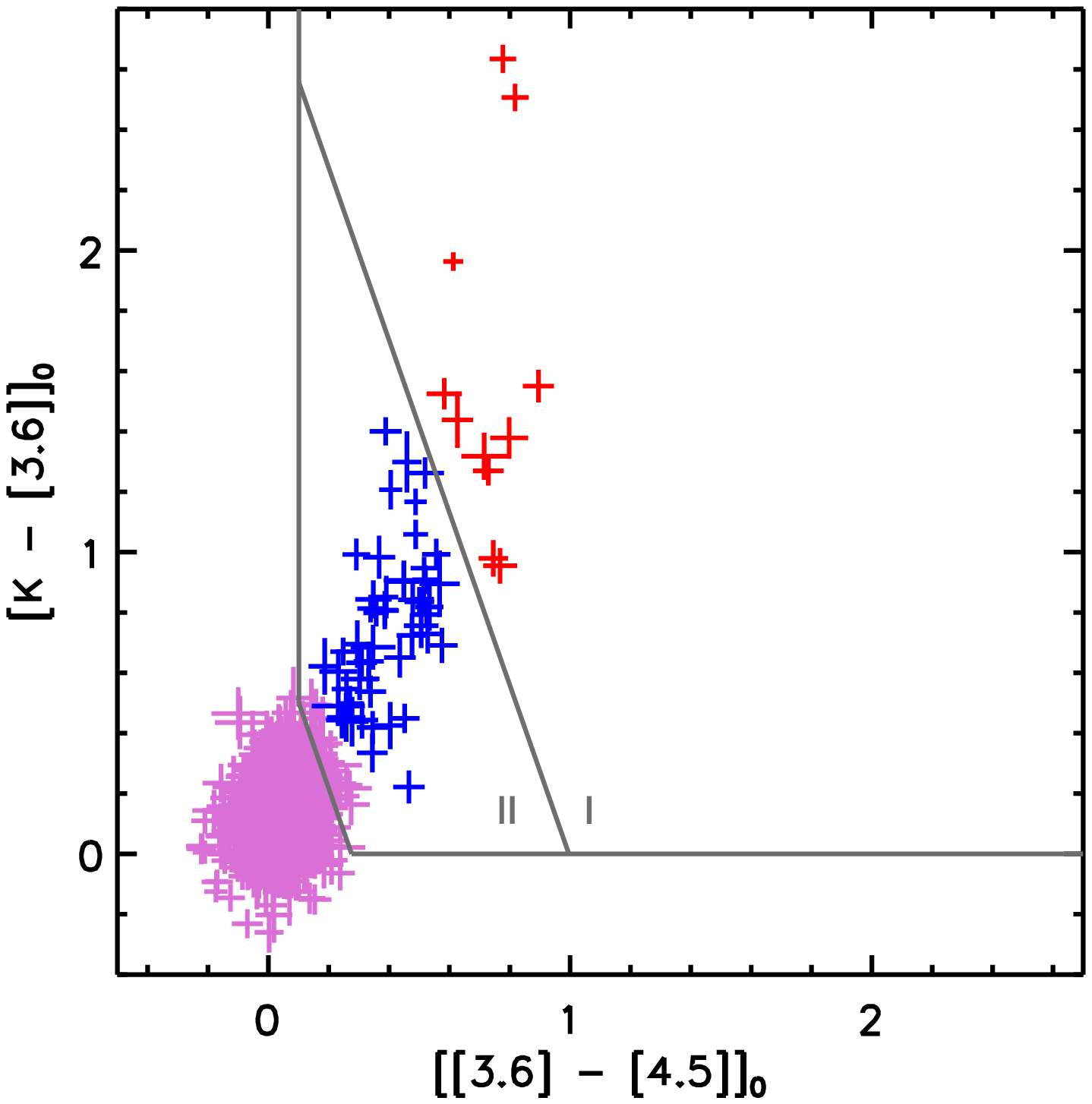}
\caption{Dereddened $K-[3.6]$ versus $[3.6]-[4.5]$ colors of the GLIMPSE360 catalog objects in the vicinity of Group 00.  Only those with adequate 2MASS photometry for classification via the scheme of \citet{gut09} are shown.  Gray lines mark the regions for Class I and II objects as indicated. Purple symbols are for objects outside the YSO loci, the 52 blue symbols are for Class II candidates, and the 11 red symbols are for Class I candidates.  The sizes of the symbols indicate the uncertainties. \label{f.s2m}}
\end{figure}

Figure~\ref{f.s2m} shows the results of this process for sources with $107.21^\circ<{\rm RA}<107.65^\circ$ and $-10.73^\circ<{\rm Dec}<-10.29^\circ$.  All sources are included that have adequately precise 2MASS photometry for classification. There are 11 Class I sources, 52 Class II sources, and 1197 sources with colors outside the YSO loci.  Table~\ref{t.s2m} compares the classification of sources in Group 00 by WISE and S2M; below we discuss the table entries in more detail.

\subsubsection{Class I Comparison}

While WISE detects 29 Class I sources in Group 00, S2M identifies only 11.  Four are common to both sets.  The seven S2M Class I sources not identified as such by WISE are a heterogeneous group.  Two are WISE Class II sources.  The other five are not identified as YSOs: One falls in the region of Figure~\ref{f.cc_annot} labeled ``shock emission,'' three are in the Class I locus but are rejected due to contamination or {\em xscprox} flags, and one is undetected by WISE, likely due to confusion with a nearby source.

We conclude that the S2M method detects between one and four Class I candidates missed by WISE: the one that is undetected by WISE due to confusion plus up to three that are in the WISE Class I locus but rejected due to flags.  Thus, WISE detects 29 of between 30 and 33 Class I sources in the combined sample, or 88 to 97\% of them.

\begin{deluxetable}{lcccc}
\tablecaption{WISE and Spitzer+2MASS (S2M) YSO Counts in Group 00\label{t.s2m}}
\tablewidth{\hsize}
\tablehead{\colhead{} & \colhead{S2M} & \colhead{S2M} & \colhead{Not a S2M} & \colhead{Total} \\
\colhead{} & \colhead{Class I} & \colhead{Class II} & \colhead{YSO} & \colhead{}}
\startdata
WISE Class I  & 4 & 0 & 25 & 29 \\
WISE Class II & 2 & 7 & 3 & 12 \\
Not a WISE YSO & 5 & 39 & \nodata & \nodata \\
Total & 11 & 46\tablenotemark{1} & \nodata & \nodata
\enddata
\tablenotetext{1}{Six sources are excluded that have the requisite colors but are further than 330\arcsec\ from Group 00 members.} 
\end{deluxetable}

Of the 25 WISE Class I candidates not classified as such by S2M, one is in the middle of two S2M sources separated by 10\arcsec, while the other 24 appear upon visual inspection to be point sources or, in three cases, close blends.  Due to insufficient or insufficiently precise 2MASS or Spitzer photometry, these are not counted as Class I sources by the \citet{gut09} method, but the existing S2M colors are not inconsistent with those of YSOs; see Section 5.1.3.  If we allow 2MASS photometry with uncertainties up to 0.2 mag instead of 0.1 mag, the number of the 29 WISE Class I sources classified as such by S2M increases from four to nine.  

\subsubsection{Class II Comparison}

While our WISE criteria identify 12 Class II sources in Group 00, the S2M criteria identify 52 Class II sources in the search region.  Because the search box extends beyond the WISE-identified objects, some S2M Class II candidates on the outskirts of the search box are likely not cluster members.  We recalculated the MST for the combined WISE and S2M samples of both classes and excluded the six S2M candidates (all Class II) that are separated from the cluster by more than 330\arcsec.  This leaves 46 S2M Class II candidates.

Seven Class II sources are common to both sets, while 39 S2M Class II sources are not identified as such by WISE.  Five of the 39 are outside the WISE YSO loci, with $W1-W2$ colors typical of Class II or borderline Class I/II sources but $W2-W3$ colors that are too small.  Two of these pass our point source criteria, while the other three have contaminant flags.  Four of the 39 are not in the AllWISE catalog; visual inspection suggests this is due to blending with nearby, brighter point sources.  The other 30 of the 39 are in the WISE Class II color space.  Of these, two have {\em xscprox} flags, two have contaminant flags, 14 fail the magnitude criteria, and 12 fail the magnitude criteria {\em and} have contaminant flags.

The large number of sources that have Class II colors in both sets but are too faint for our WISE criteria suggests that an iterative method, where the magnitude cutoffs designed to exclude extragalactic sources are relaxed in the vicinity of an identified YSO cluster, may be useful in recovering more of the YSO population with WISE.  Indeed, relaxing the $W1$ cutoff to 13 mag instead of 12 mag increases the number of sources common to both Class II catalogs from 7 to 18.  We leave this approach for future work.

In Group 00, we conclude that S2M detects at least 18 Class~II candidates missed by WISE: the four that are not in the AllWISE catalog due to blending plus 14 that are in the WISE Class II locus but fail the magnitude criteria.  Up to 16 additional sources that are in the WISE Class II locus but are rejected due to flags may also belong on this list.  Thus, WISE detects 12 of between 30 and 46 Class II sources in the combined sample, or 26 to 40\% of them.

Of the five WISE Class II candidates not classified as such by S2M, two have Class I colors.  Of the remaining three, one is outside the YSO color spaces, and two are in Class II color space but lack sufficiently precise 2MASS data to be classified as such.

\subsubsection{Summary}

We conclude that, in Group 00, WISE colors are sufficient to identify nearly all the Class I sources accessible to the combined powers of WISE, post-cryogenic Spitzer, and 2MASS, while WISE identifies fewer than half of the Class II sources in such a sample.  This is consistent with the expectations set forth in Section 3.3 based on the models of \citet{bar15}.  Of the two methods considered, and at a distance of 1000 pc, the combination of GLIMPSE360 and 2MASS is best for finding Class II sources, while WISE is best for finding Class I sources.

\begin{figure}
\includegraphics[width=\hsize,trim={0 0 0 32.5},clip]{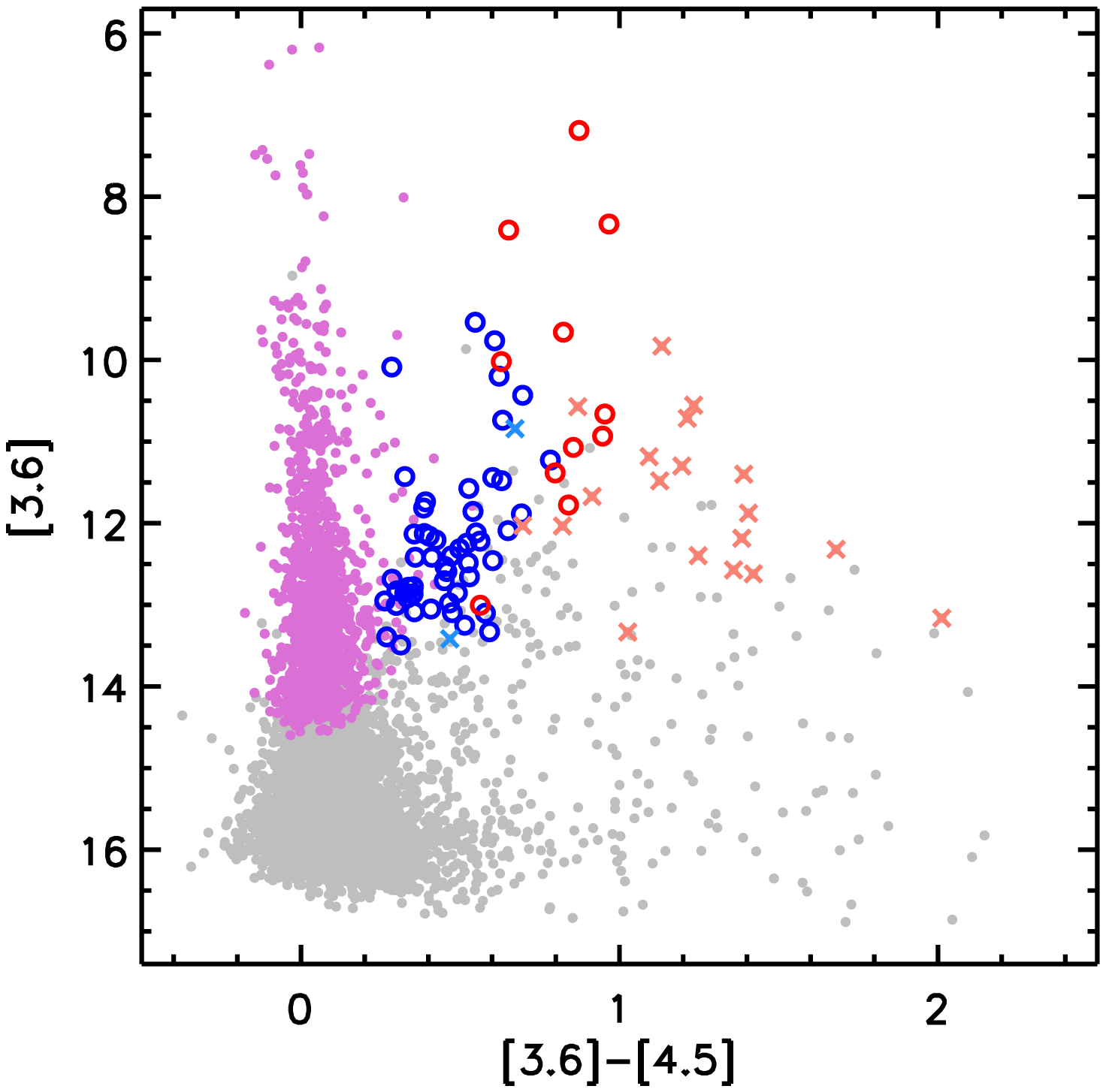}
\caption{Observed 3.6 \micron\ magnitude versus the $[3.6]-[4.5]$ color of all sources in the vicinity of Group 00 with Spitzer uncertainties $<0.1$ mag and without regard to the quality of the 2MASS detection.  Sources classified in Figure~\ref{f.s2m} are shown with the same color scheme as before.  Sources with insufficent 2MASS photometry for classification via the \citet{gut09} method are shown with salmon $\times$ symbols if they are Class I via WISE, light blue $\times$ symbols if they are Class II via WISE, and gray if they are not classified as YSOs by our WISE criteria. \label{f.glimpse}}
\end{figure}

In Figure~\ref{f.glimpse}, we consider the potential to identify Class~I YSOs with post-cryogenic Spitzer data alone.  It plots the GLIMPSE360 3.6 \micron\ magnitude versus the $[3.6]-[4.5]$ color of all sources near Group 00 with Spitzer uncertainties $<0.1$ mag, regardless of the quality of the 2MASS detection. Magenta, red, and blue symbols mark the sources previously classified in Figure~\ref{f.s2m}. Crosses mark sources that lack sufficiently precise 2MASS data for classification by the \citet{gut09} method but are placed in Class I (salmon) or Class II (light blue) by WISE.  While there are few such Class II sources, there are many Class I sources.  Without supplemental data from other surveys, two-band post-cryogenic Spitzer data are insufficient to identify Class I sources unambiguously, but this analysis suggests that sources in known star-forming regions with $[3.6]<12.5$ mag and $[3.6]-[4.5]>1$ mag are good candidates.

\subsection{Calibrating WISE Flat-Spectrum and Class 0 Colors with the Herschel Orion Protostar Survey}

Our initial WISE selection criteria do not identify flat-spectrum or Class 0 protostars.  Flat-spectrum sources are known as such due to their near-zero spectral slopes, intermediate to those of Class I and Class II objects.   Identifying these for follow-up study is important, since their nature is not entirely clear \citep{dun14}.  They may represent an evolutionarily homogeneous transition from envelope-embedded to disk-dominated systems, or they may represent systems with a range of evolutionary states but combinations of inclination, foreground reddening, and envelope density that yield flat SEDs \citep{fur16}.

The Class 0 protostars are defined in part by their far-IR to submillimeter SEDs, in which more than 0.5\% of the source luminosity is emitted beyond 350 \micron\ \citep{and93}.  This observational class is expected to correspond roughly to a physical stage where more than half of the final mass of the star still resides in the envelope \citep{dun14}.  With photometry only at wavelengths $\le 22$ \micron, it is expected to be difficult to detect Class 0 protostars in WISE and to distinguish them from Class I protostars, but placing a lower limit on the number of especially young protostars outside the well known molecular clouds would constrain the importance of isolated modes of star formation.

To estimate the number of flat-spectrum and Class 0 protostars among our Canis Major candidates, we used the well characterized protostars of HOPS, the Herschel Orion Protostar Survey \citep{stu13,man13,fur16}, to explore where Class 0, Class I, and flat-spectrum protostars lie in WISE color-color diagrams.  HOPS was a Herschel key program that acquired 50--200 \micron\ spectroscopy and 70 and 160 \micron\ imaging of Spitzer-identified protostars in the Orion A and B molecular clouds, home to the largest population of protostars in the nearest 500 pc \citep{meg12}.  As such, the HOPS sample is ideal for calibrating the expected characteristics of protostellar candidates in other regions.

After removing likely extragalactic contaminants and objects with poorly sampled far-IR photometry, \citet{fur16} selected 330 HOPS targets for further study.  They constructed the 1 to 870 \micron\ SEDs of the targets with photometry and spectra from 2MASS, Spitzer, Herschel, and the Atacama Pathfinder Experiment (APEX).  Using bolometric temperatures and mid-IR SED slopes, they identified 92 Class~0 protostars, 125 Class I protostars, 102 flat-spectrum protostars, and 11 Class II objects in this sample.

To get WISE photometry for these sources, we searched AllWISE for the closest match to each Spitzer position.  For 236 of 330 sources, there was a match within 1\arcsec; these were automatically paired with their Spitzer counterparts.  For 32 sources, there was no match within 10\arcsec; these were automatically rejected from further analysis.  We manually inspected the remaining 62 sources at intermediate separations.  In these cases, the larger offset was due either to scattered light from the Spitzer point source or to a blend in WISE of two or more distinct Spitzer sources.  We retained the 46 matches where the bulk of the WISE flux was judged to be due to the Spitzer source in question (all with separations less than 7\arcsec) and rejected the other 16.

\begin{figure}
\includegraphics[width=\hsize,trim={0 0 0 32.5},clip]{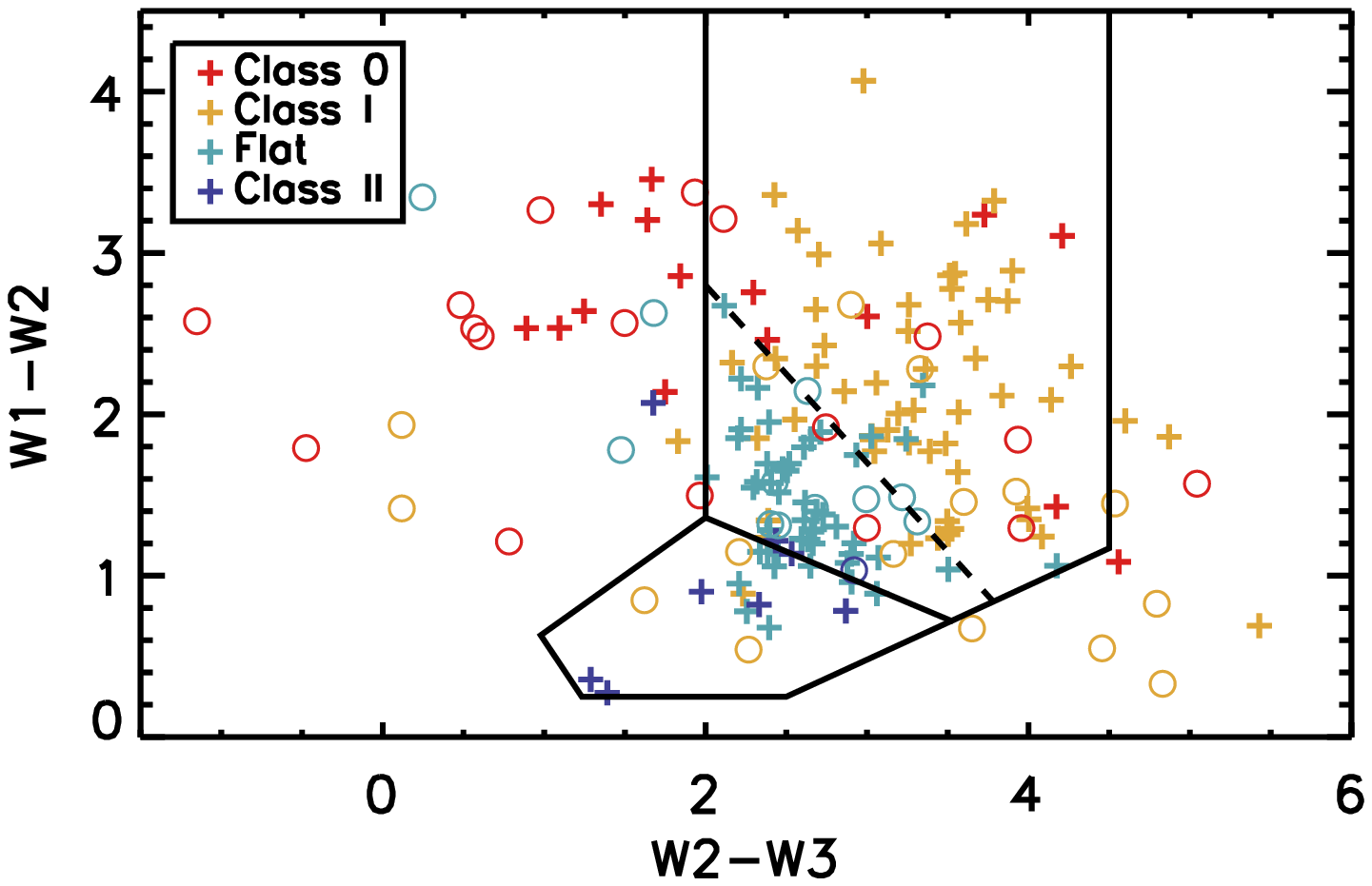}
\caption{$W1-W2$ versus $W2-W3$ colors of WISE counterparts to 173 HOPS sources that pass the point source criteria.  As indicated in the legend, symbol color indicates the YSO class determined from analysis of 2MASS, Spitzer, Herschel, and APEX data \citep{fur16}.  Sources marked with $+$ signs satisfy our magnitude cuts, while those marked with $\circ$ signs do not.  Solid black lines mark the regions for WISE Class I (upper) and Class II (lower) candidates.  The dashed line marks our proposed boundary between flat-spectrum and Class I YSOs.\label{f.hopswise}}
\end{figure}

Of the 282 Spitzer sources with WISE counterparts, 173 survived the WISE point-source quality cuts described in Section 3.1; i.e., they were detected in Bands 1 and 2 and not flagged as artifacts or extended sources.  Nearly all of the 109 rejected sources had extended emission or Band 3 or 4 artifact flags, likely due to the copious background in some parts of Orion.  The 173 valid matches include 32 of 92 Class 0 sources (35\%), 72 of 125 Class I sources (58\%), 60 of 102 flat-spectrum sources (59\%), and 9 of 11 Class II sources (81\%).  The increase in the percentage from Class 0 to Class~I and flat-spectrum sources supports our earlier conjecture that WISE is more effective in detecting more evolved protostars.  Figure~\ref{f.hopswise} shows these 173 HOPS sources on a WISE $W1-W2$ versus $W2-W3$ diagram, and Table~\ref{t.hops} shows how the HOPS classifications correspond to the WISE classifications.

\begin{deluxetable}{lc@{}cc@{}cc@{}cc@{}c}
\tablecaption{WISE Classification of Herschel Orion Protostars\tablenotemark{1} \label{t.hops}}
\tablewidth{\hsize}
\tablehead{\colhead{} & \multicolumn{8}{c}{HOPS Class} \\ \cline{2-9} \\ [-1.5ex]
\colhead{} & \multicolumn{2}{c}{0} & \multicolumn{2}{c}{I} & \multicolumn{2}{c}{Flat} & \multicolumn{2}{c}{II}}
\startdata
WISE Class I  & 6 & (6)  & 51 & (6)  & 39 & (8)  & 1 & (1) \\
WISE Class II                   & 0 & (0)  & 1  & (3)  & 10 & (0)  & 6 & (0) \\
$W2-W3$ too small\tablenotemark{2}    & 8 & (10) & 1  & (2)  & 0  & (3)  & 1 & (0) \\
$W2-W3$ too large\tablenotemark{2}   & 1 & (1)  & 3  & (5)  & 0  & (0)  & 0 & (0) \\
Fails Point Source Criteria & \multicolumn{2}{c}{29}    & \multicolumn{2}{c}{42}     & \multicolumn{2}{c}{36}     & \multicolumn{2}{c}{2}     \\
No AllWISE Counterpart      & \multicolumn{2}{c}{31}    & \multicolumn{2}{c}{11}     & \multicolumn{2}{c}{6}      & \multicolumn{2}{c}{0}     \\
\hline \\ [-1.5ex]
Total        & \multicolumn{2}{c}{92}    & \multicolumn{2}{c}{125}    & \multicolumn{2}{c}{102}    & \multicolumn{2}{c}{11}
\enddata
\tablenotetext{1}{The first four rows tabulate HOPS sources with valid WISE counterparts. Counts outside parentheses are of sources that satisfy the WISE magnitude criteria, while counts inside parentheses are of those that do not.} 
\tablenotetext{2}{These would not be classified as YSOs by our WISE-only criteria, as they fall outside the requisite parts of the color-color space.}
\end{deluxetable}

In general, the HOPS classifications, which are based on the full 1 to 870 \micron\ SEDs, agree well with the WISE classifications based only on 3.4 to 22 \micron\ photometry.  Of the 173 WISE counterparts that pass the point source criteria, the majority are identified as YSOs in WISE, where 138 (80\%) lie in the WISE Class I or Class II loci, and 114 (66\%) lie in these loci and pass the magnitude tests.  Comparing classes directly and ignoring magnitude cuts, 80\% of the HOPS sources in the WISE Class II locus are Class II or flat-spectrum sources, and 98\% of the HOPS sources in the WISE Class I locus are flat-spectrum, Class I, or Class 0 protostars.

\subsubsection{Flat-Spectrum Sources}

Within the WISE Class I locus, the flat-spectrum HOPS sources (light blue symbols in Figure~\ref{f.hopswise}) are concentrated toward smaller colors.  We suggest a division in the $W1-W2$ versus $W2-W3$ color space between flat-spectrum and Class~I protostars.  This is shown as the dashed line in Figure~\ref{f.hopswise}; flat-spectrum sources typically have \begin{equation}W1-W2 < -1.1\times(W2-W3)+5.\end{equation}

Including only sources that satisfy the magnitude cuts and would thus be counted as WISE YSOs, 51 of the 56 HOPS sources above this line in the Class I locus (91\%) are Class 0 or I, and the rest are flat-spectrum.  Of the 41 HOPS sources below this line in the Class I locus, 34 (83\%) are flat-spectrum, and the rest are Class 0, I, or II.  Of the 144 WISE Class I candidates in Canis Major, 80 (56\%) are likely flat-spectrum protostars according to Equation 10.

\subsubsection{Class 0 Sources}

\begin{figure}
\includegraphics[width=\hsize,trim={0 0 0 32.5},clip]{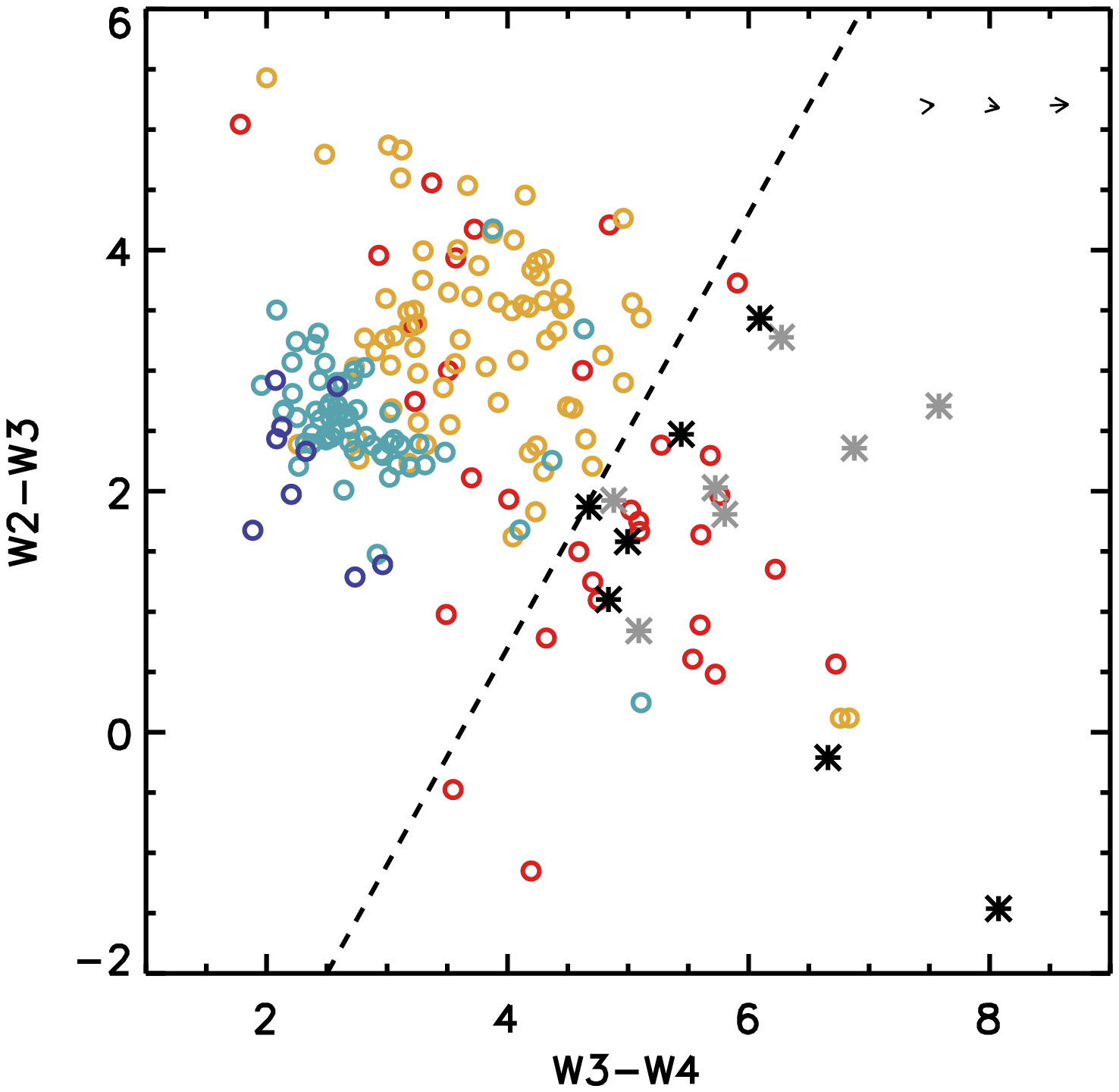}
\caption{Circles show $W2-W3$ versus $W3-W4$ colors of WISE counterparts to 173 HOPS sources that pass the point source criteria. As in the previous figure, symbol color indicates the class determined for each object by \citet{fur16}. HOPS sources below and to the right of the dashed line are almost exclusively of Class 0.  Asterisks show 14 WISE sources in Canis Major that have colors consistent with our proposed Class 0 criteria.  Black symbols represent the seven most likely Class 0 candidates, while gray symbols are likely to have their colors affected by nearby sources that are bright in $W4$.  The tiny arrows in the upper right corner of the plot indicate reddening vectors for, from left to right, $A_K=0.4$, 0.8, and 2.\label{f.w2w3w4}}
\end{figure}

In Figure~\ref{f.hopswise}, the HOPS Class 0 protostars (red symbols) show a range of $W2-W3$ colors; the majority lie outside the Class I and Class II color spaces.  Scattered light from outflow cavities or shock-excited H$_2$ emission may enhance the $W2$ fluxes \citep{tob07}, and deep 10 \micron\ silicate features in typical Class 0 sources may extinguish the $W3$ fluxes relative to those of Class I sources.  Both of these effects would yield small or possibly negative $W2-W3$ colors.

In the $W1-W2$ versus $W2-W3$ color space, almost all HOPS Class 0 sources either overlap with HOPS Class I YSOs or lie in a part of color-color space typical of shock emission.  To better identify Class 0 candidates in Canis Major, we instead look at the $W2-W3$ versus $W3-W4$ color space in Figure~\ref{f.w2w3w4}.  Here we can draw a line between a region consisting predominantly of Class I, flat, and Class II sources and a region consisting predominantly of Class 0 sources.  The Class 0 region is described as \begin{equation}W2-W3 < 1.8\times(W3-W4)-6.5.\end{equation}  In this region, 19 of 22 HOPS targets (86\%) are of Class 0.  On the other side of the line, only 13 of 151 HOPS targets (9\%) are of Class 0.

We add one more requirement, motivated by the location of the HOPS Class 0 sources in Figure~\ref{f.hopswise}, that WISE Class 0 candidates have \begin{equation}W1-W2 > 1.\end{equation}  The fourteen sources in Canis Major that meet all point source criteria, the magnitude criteria discussed above, and the two criteria in this section are plotted as asterisks in Figure~\ref{f.w2w3w4}.  These were not necessarily identified as YSO candidates in previous steps.

A potential problem with using $W3-W4$ colors in the selection criteria is that, due to the larger $W4$ beam, sources with clean detections in the shortest three bands can have artificially elevated $W4$ photometry due to contamination by a nearby bright source.  To exclude catalog detections with inflated band 4 photometry, we visually inspected $W4$ images of all 14 candidates, rejecting those where there was no obvious point source in band 4, but there was one nearby.  The rejected sources were all within 48\arcsec\ of another source with at least twice as much $W4$ flux, while the accepted sources had no such bright sources within this radius.  This would be an appropriate criterion for automatic inspection of $W4$ sources. Seven of the 14 candidates, shown in black in Figure~\ref{f.w2w3w4}, survived this process.

\begin{figure}
\includegraphics[width=\hsize]{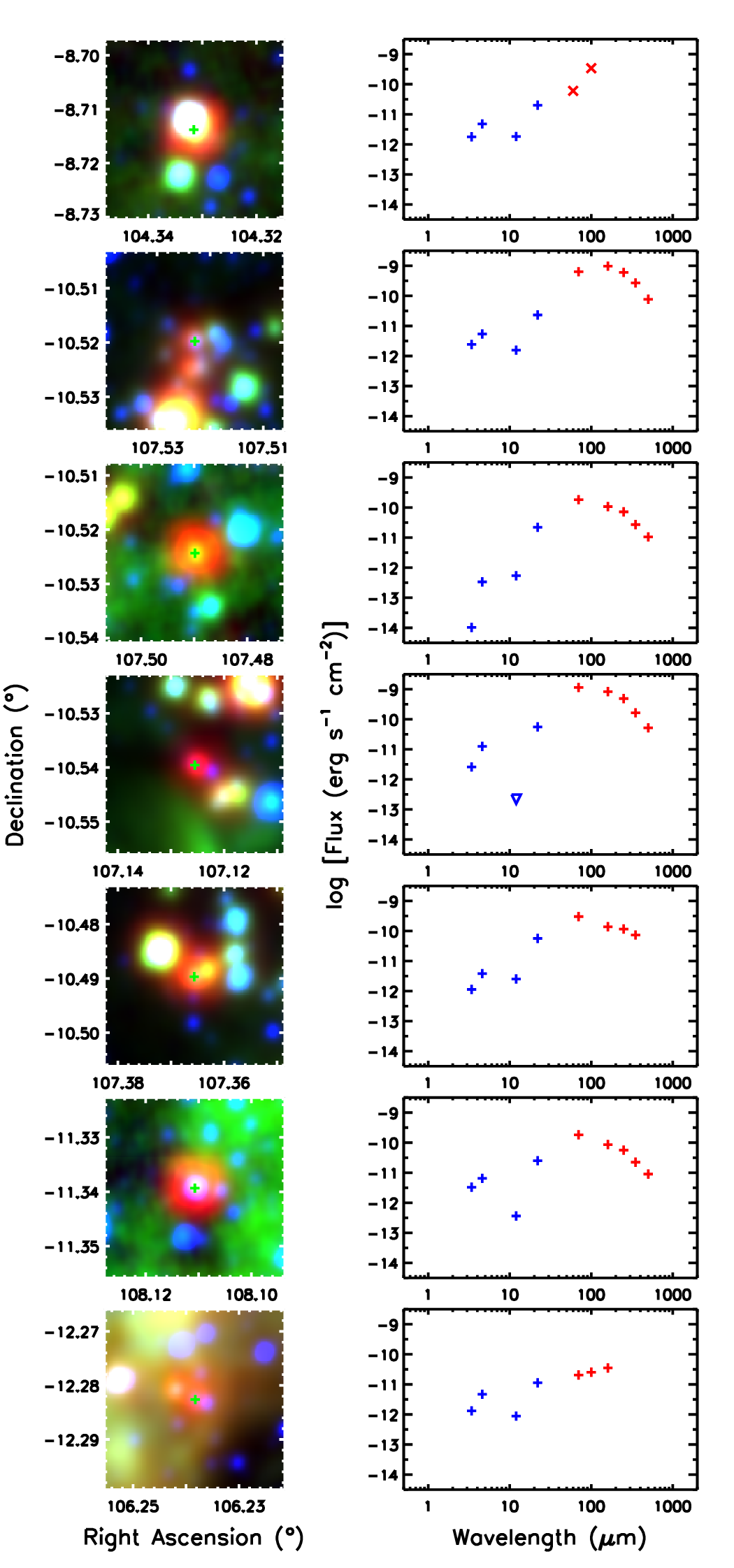}
\caption{{\em Left:} WISE images of the Class 0 candidates marked with black symbols in Figure~\ref{f.w2w3w4}. Blue is 3.4 \micron, green is 12 \micron, and red is 22 \micron.  The stretch in each band is from the minimum to the 95th percentile.  Their AllWISE positions are indicated by $+$ signs. {\em Right:} SEDs of the sources. Blue symbols are for WISE data ($\nabla$:\ upper limit); red symbols are for Herschel ($+$) or IRAS ($\times$) data. Candidates are presented in the order given in Table~\ref{t.class0}.\label{f.class0}}
\end{figure}

\begin{deluxetable*}{lcccrcccccc}
\tablecaption{Class 0 Candidates in CMa\tablenotemark{1}\label{t.class0}}
\tablewidth{\hsize}
\tablehead{\colhead{WISE ID} & \colhead{Other ID} & \colhead{Group} & \colhead{RA (deg)} & \colhead{Dec (deg)} & \colhead{$W1/W2/W3$} & \colhead{$L_{\rm bol}$} & \colhead{$L_{\rm smm}$} & \colhead{$L_{\rm smm}/L_{\rm bol}$} & \colhead{$T_{\rm bol}$} & \colhead{$T_{\rm bol}$} \\ \colhead{} & \colhead{} & \colhead{} & \colhead{} & \colhead{} & \colhead{Class} & \colhead{($L_\odot$)} & \colhead{($L_\odot$)} & \colhead{Class} &\colhead{(K)} & \colhead{Class}}
\startdata
J065719.54$-$084249.1 & IRAS 06459$-$0838 & \nodata & 104.33143 & $-$8.71367  & \nodata & 4.2  & \nodata & \nodata  & 68.6 & 0 \\
J071005.46$-$103110.6 & \nodata           & 00      & 107.52275 & $-$10.51964 & \nodata & 48.4 & 2.2     & 0        & 37.1 & 0 \\
J070957.52$-$103127.1 & \nodata           & 00      & 107.48969 & $-$10.52421 & I       & 9.6  & 0.2     & 0        & 48.2 & 0 \\
J070830.08$-$103221.9 & \nodata           & 02      & 107.12535 & $-$10.53942 & \nodata & 60.1 & 1.3     & 0        & 45.2 & 0 \\
J070927.71$-$102922.3 & \nodata           & 00      & 107.36546 & $-$10.48954 & I       & 15.8 & 2.3     & 0        & 55.6 & 0 \\
J071226.53$-$112021.1 & \nodata           & \nodata & 108.11056 & $-$11.33920 & \nodata & 9.3  & 0.2     & 0        & 58.0 & 0 \\
J070457.14$-$121656.9 & LDN 1657A-1       & 05      & 106.23811 & $-$12.28248 & \nodata & 1.5  & \nodata & \nodata  & 99.5 & I
\enddata
\tablenotetext{1}{Luminosities assume a distance of 1000 pc.}
\end{deluxetable*}

Figure~\ref{f.class0} shows images and SEDs of the seven Class 0 candidates, and Table~\ref{t.class0} lists their AllWISE identifiers and other properties.  While four of the sources have no counterpart in SIMBAD, one source has an IRAS designation, and one, LDN 1657A-1, was identified as a star-forming core by \citet{lin10}.  Three of the candidates are in Group 00, one is in Group 02, one is in Group 05, and the other two are not connected to any group.  They are close to molecular clouds found by \citet{kim04}; they are all on or within those authors' $^{13}$CO contours at 1~K~km~s$^{-1}$.  Two of them are classified as Class I sources in the $W1-W2$ versus $W2-W3$ color space, while the other five are in the region labeled ``Shock Emission'' in Figure~\ref{f.cc_annot}.

The far-infrared flux densities in the SEDs come from three compilations.  For the five sources between declinations $-10^\circ$ and $-12^\circ$, Herschel data, obtained as part of the Hi-GAL key program \citep{mol10}, are taken from \citet{eli13}.  The photometry of LDN 1657A-1 is taken from \citet{rag12}, which reported results from the Herschel EPoS key program.  The northernmost source does not appear to have been observed by Herschel, but it is in the IRAS point source catalog.  We plot its IRAS flux densities at 60 and 100~\micron.

With the combined WISE and far-infrared data, we calculate the bolometric luminosity ($L_{\rm bol}$), submillimeter luminosity ($L_{\rm smm}$), and bolometric temperature ($T_{\rm bol}$) of each source.  These properties are less certain for sources that lack the full complement of long-wavelength photometry or that may be affected by blending with nearby sources.  The submillimeter luminosity is derived from photometry at 350 \micron\ and longer, for consistency with \citet{and00} and subsequent authors.  At least one of the 350 and 500 \micron\ points must be available to calculate this quantity.  Given these caveats, the classifications based on the full SEDs (discussed in the Introduction) are in good agreement with the WISE classifications.  All of the WISE Class 0 candidates with the requisite data are also Class 0 by the submillimeter luminosity ratio, and six of the seven candidates are Class 0 by bolometric temperature.  The other, LDN 1657A-1, is Class I, but it lies not far above the Class 0/I boundary of 70 K.  \citet{eli13} and \citet{rag12} fit modified blackbody functions to the six sources observed with Herschel. They found properties typical of cold, young YSOs that are consistent with what we obtain from direct integration of the SEDs.

Given that 34\% (31/92) of the Orion Class 0 protostars have no counterpart in AllWISE, the WISE-based criteria presented here are unlikely to select the full Class 0 population in a star-forming region. They are likely, however, to effectively identify a subsample of sources for targeted far-infrared or submillimeter follow-up.

\section{DISCUSSION}

\subsection{The Diversity of Class II to Class I Ratios in CMa OB1}

As discussed in the introduction, Canis Major contains a canonical example of star formation suspected to be induced by a supernova \citep{her77}. The ring of optical and radio emission described in that work coincides with the largest YSO groups we find in our search (Figure 22), and there are relatively few isolated YSOs, all of Class II, near the center of the ring.

If the Class II to Class I ratio is a proxy for group age, the diversity of ratios in the groups around the ring is puzzling.  Naively, one would expect the groups at a given distance from the explosion to be of similar age.  Group 00 on the north side of the ring contains the most Class I sources of any group, while Groups 04 and 05, to the southeast and southwest, are dominated by Class II sources.  The bright background due to S296 along the western side of the ring may decrease our sensitivity to Class I sources there, but this cannot be the case in the darker regions on the eastern side.

Variables other than age may play a role in determining the Class II to Class I ratio.  We find that, around the supernova remnant, the Class II to Class I ratio is inversely correlated with the cloud masses tabulated by \citet{kim04}.  The clouds on the north of the ring, where there are five groups with $N_{\rm II}/N_{\rm I}<2$, have masses of $1.2\times10^4$ $M_\sun$ and $1.6\times10^4$~$M_\sun$.  The cloud to the southeast is an intermediate case, where the cloud mass is 7500 $M_\sun$ and $N_{\rm II}/N_{\rm I}$ for the associated groups are 4.0 and 5.0.  The clouds to the southwest have masses less than 1000 $M_\sun$, and the two groups there have $N_{\rm II}/N_{\rm I}$ of 8.0 and 8.3.  These exact results depend on the assumptions we made in dividing the YSOs into groups (Section 4), but the broad trend of larger Class II ratios associated with less massive clouds will hold for other clustering schemes.

It is unclear whether such a trend exists for groups and clouds outside CMa OB1 and the supernova remnant, due to the small number of YSO groups that are associated with clouds in G220.8$-$1.7 and the other complexes.  We note that the BFS 64 cloud, which appears to host our Group 06, is tentatively placed at 3.2 kpc by \citet{kim04}, a factor of three larger than other clouds in the search field, although they discuss ambiguities in this result.  The published mass of the cloud and the fraction of each YSO class recovered by WISE depend on distance, so these quantities have additional uncertainties in this part of the search region.

\subsection{Comparison of the Canis Major Groups to Other Star-Forming Regions\label{s.comp}}

To place the groups identified in Section 4 in the context of others that have been studied in depth, we compare their properties to those tabulated by \citet{gut09} in their Spitzer survey of 36 young stellar clusters within 1 kpc of the Sun.  \citeauthor{gut09}\ characterize both clusters and cores within the clusters.  The projected areas and source separations of our groups more closely resemble those of the clusters, not the cores, so we make comparisons with the former.

Table~\ref{t.clusmed} shows the median YSO, Class I, and Class II counts, the median Class II to Class I ratio, and the median area, YSO surface density, and nearest-neighbor distance for the Canis Major groups and the \citet{gut09} clusters.  It also tabulates the Kolmogorov-Smirnov (KS) probability that the Canis Major and \citeauthor{gut09} samples are drawn from the same underlying distribution.  There are, broadly speaking, fewer detected members separated by greater distances in the Canis Major groups than in major clusters within 1 kpc.

The fourth through seventh columns of Table~\ref{t.clusmed} address how the group properties change if, due to the incompleteness of WISE to Class II YSOs at 1 kpc (Sections 3.3 and 5.1), we assume that the actual population of Class II YSOs in each group is twice or four times that detected.  The median Class~I counts are obviously unchanged, the median Class II counts double or quadruple as expected, and the median total YSO counts increase by 86\% and 271\%.  The median ratios of Class~II to Class I YSOs double and quadruple, as expected.

\begin{deluxetable*}{lccccccc}
\tablecaption{Median Properties of Groups with More Than Four Members\label{t.clusmed}}
\tablewidth{\hsize}
\tablehead{\colhead{Property} & \multicolumn{2}{c}{Raw} & \multicolumn{2}{c}{$N_{\rm II}$ Doubled\tablenotemark{1}} & \multicolumn{2}{c}{$N_{\rm II}$ Quadrupled\tablenotemark{1}} & \colhead{G09\tablenotemark{2}} \\ 
\cline{2-3} \cline{4-5} \cline{6-7} \\ [-1.5ex]
\colhead{} & \colhead{CMa} & \colhead{$\log ({\rm KS~Prob.})$\tablenotemark{3}} & \colhead{CMa} & \colhead{$\log ({\rm KS~Prob.})$\tablenotemark{3}} & \colhead{CMa} & \colhead{$\log ({\rm KS~Prob.})$\tablenotemark{3}} & \colhead{}}
\startdata
$N_{\rm YSO}$          & 14   & $-$4.5 & 26   & $-$1.8 & 52   & $-$0.6 & 49    \\ 
$N_{\rm I}$            & 3    & $-$2.0 & 3    & $-$2.0 & 3    & $-$2.0 & 10    \\ 
$N_{\rm II}$           & 12   & $-$5.5 & 24   & $-$2.3 & 48   & $-$0.5 & 41    \\ 
$N_{\rm II}/N_{\rm I}$ & 1.6  & $-$2.0 & 3.1  & $-$0.4 & 6.2  & $-$1.4 & 3.8   \\ 
$A$ (pc$^2$)           & 11.  & $-$1.8 & 11.  & $-$1.3 & 11.  & $-$1.6 & 5.3   \\ 
$\sigma$ (pc$^{-2}$)   & 1.8  & $-$5.7 & 2.9  & $-$4.4 & 4.8  & $-$2.6 & 13    \\ 
NN2$_{\rm med}$ (pc)   & 0.28 & $-$5.8 & 0.21 & $-$4.5 & 0.16 & $-$2.2 & 0.089
\enddata
\tablenotetext{1}{Results if each CMa group contains a population of randomly distributed, undetected Class II YSOs such that $N_{\rm II}$ is two or four times that observed.}
\tablenotetext{2}{Median properties of the 36 clusters in \citet{gut09}.}
\tablenotetext{3}{Logarithm of the probability that the CMa properties are drawn from the same underlying distribution as the \citet{gut09} properties.}
\end{deluxetable*}

To check the effect of these corrections on the group area, surface density of YSOs, and the median separation of YSOs, we examined randomly generated groups.  For each observed group, we generated 1000 groups with YSOs equal in number to the observed one but randomly distributed in space, calculated their areas and median separations, and then doubled and quadrupled the number of Class II sources.  These new sources were randomly distributed in a circle centered at the mean group position and having the same area as the original group.  The areas remained approximately the same, changing by up to 10\% in the smallest groups due to significant changes in the number and location of vertices.  The median separations decreased by up to 32\% in the doubling case and by up to 54\% in the quadrupling case, with the largest fractional decreases for the groups with the largest Class II to Class I ratios.

With these corrections for missing Class II YSOs, the 16 Canis Major groups remain less populated and more diffuse than the 36 nearby \citet{gut09} clusters.  However, some of the KS probabilities exceed 10\%, namely the $N_{\rm II}/N_{\rm I}$ ratio in the doubling case and the YSO and Class II counts in the quadrupling case.  This suggests that the undercounting of Class II YSOs by WISE at 1 kpc is partially responsible for the observed differences from nearby clusters.

We performed a similar comparison to the Orion groups and clusters tabulated by \citet{meg16}.  We considered the properties of 21 Orion groups and clusters after a correction for incompleteness, and we compared them to our 16 groups of more than four members after the correction for missing Class II YSOs.  Three of the Orion clusters (the Orion Nebula Cluster, NGC 2024, and NGC 2068/2071) have hundreds to thousands of YSOs and are much larger than anything in Canis Major.

With the doubled Class II counts in Canis Major, the remaining, smaller groups and clusters in Orion have YSO counts of similar magnitude to those of the Canis Major groups.  The median Orion group has 37 members versus 26 members for Canis Major (KS probability 0.11).  The Orion groups are more compact, however, with median area and density of 1.8 pc$^2$ and 21 YSO pc$^{-2}$ compared to 11 pc$^2$ and 2.9 YSO pc$^{-2}$ for Canis Major.  The KS probabilities that the areas and densities of the groups are drawn from the same distributions are $3.7\times10^{-3}$ and $4.2\times10^{-7}$.    With the quadrupled Class II counts in Canis Major, the median group has 52 members (KS probability 0.78), and the median area and density are 11 pc$^2$ and 4.8 YSO pc$^{-2}$ (KS probabilities $1.2\times10^{-3}$ and $3.5\times10^{-6}$).  Different YSO and group identification techniques clearly need to be accounted for in this comparison, but a thorough analysis is beyond the scope of this discussion.

\citet{smi10} looked at the clustering properties in Carina, at 2.3 kpc.  They identified 16 clusters in the central part of this massive star forming region with between six and 104 YSOs.  The median number of YSOs is 16.  Except for the largest cluster (the Treasure Chest), the distribution of cluster sizes is similar to that of Canis Major.  The median number of YSOs in the Canis Major clusters before the correction for missing Class II sources is 14, and the KS probability that the cluster sizes in Carina and Canis Major are drawn from the same distribution is 0.91.  As in Canis Major, the Carina counts are affected by the insensitivity of the respective detectors to lower-mass YSOs at the distance of the region.

\section{CONCLUSIONS}

With the AllWISE database, we used color and magnitude cuts to identify 479 YSOs in a $10^\circ \times 10^\circ$ region centered on the Canis Major star-forming region.  These are divided into Class I and Class II YSOs, of which there are 144 and 335, respectively.  At the adopted distance of 1000 pc, the criteria are sensitive to Class I YSOs down to $\sim0.2$ $M_\sun$ and to Class II YSOs, with their fainter mid-IR photometry, down to $\sim0.5$~$M_\sun$.  Thus, we expect the number of Class II non-detections to be significant.  Applying the search criteria to adjacent regions that are expected to be devoid of star formation, we estimate that 11\% of our Class I candidates and 16\% of our Class II candidates may be non-YSO point source contaminants.

We calculated the minimum spanning tree of the YSO distribution and concluded that there are 16 groups with more than four members.  Of the 479 YSOs, 53\% are in such groups.  The groups have a wide range of Class II to Class~I ratios.  These range from 0.4 to 8.3 among groups with more than 25 members and extend to more extreme values among the smaller groups.  The groups are generally aligned with the locations of $^{13}$CO clouds mapped by \citet{kim04}, and more massive clouds tend to be aligned with more heavily populated groups with small Class II to Class I ratios.  In CMa OB1, the groups with the smallest ratios are roughly concentrated along the northern edge of the supernova remnant, while those with larger ratios are in the southeast and southwest.   

We examined the results from our WISE selection process in light of data sets from 2MASS, Spitzer, and Herschel.  At the distance of Canis Major, WISE recovers essentially all of the Class I population that is accessible to WISE, 2MASS, and post-cryogenic Spitzer.  The latter two facilities are useful in recovering the Class II population more completely than WISE can, alone.  Using the well characterized Orion protostars as a training set, we determined where flat-spectrum and Class 0 protostars lie in WISE color-color spaces.  Sources identified as WISE protostars that have $W1-W2<-1.1\times(W2-W3)+5$ are likely to be flat-spectrum sources.  Eighty of the 144 Class I candidates fall in this flat-spectrum locus.

In the $W1-W2$ versus $W2-W3$ diagram, the majority of the Orion Class 0 protostars have similar $W1-W2$ colors but bluer $W2-W3$ colors than the \citet{koe14} Class I locus.  The $W2-W3$ versus $W3-W4$ diagram coupled with a manual or automated check for $W4$ contamination is more effective in separating Class I and Class 0 protostars.  Sources with $W2-W3<1.8\times(W3-W4)-6.5$ are likely to be of Class 0.  Seven of the Canis Major sources in this range appear to be Class 0 protostars suitable for follow-up at higher angular resolution.

The distribution of YSOs and groups in the central part of the search field is consistent with the claim by \citet{her77} that star formation there was induced by a supernova; however, the range in Class II to Class~I ratios of these groups is surprising if they are all of the same age.  In this region, the Class II to Class I ratio is inversely correlated with the cloud mass measured by \citet{kim04}, suggesting that initial conditions may be an important factor in the ratio distribution.

The groups are less populous and more diffuse than those characterized by \citet{gut09} in the nearest 1 kpc; this may be in part due to undercounting with this combination of angular resolution, sensitivity, and distance.  If we assume that only half the Class II population is detected, the Class II to Class I ratios are consistent with being drawn from the same underlying distribution as the \citeauthor{gut09}\ sample.  If we assume that only one quarter of the Class II population is detected, the YSO and Class~II counts would closely resemble those of the \citeauthor{gut09}\ clusters.  The groups have similar membership counts to the smaller groups and clusters in Orion \citep{meg16} and Carina \citep{smi10}, but they are more spread out than the Orion groups.

Optical and near-infrared spectra of approximately 40 candidates in four of the largest groups have recently been acquired. In a future publication, we will use these spectra to confirm or reject the YSO classifications and, when possible, to estimate spectral types and accretion rates.

The WISE mid-IR all-sky survey is effective in uncovering instances of low-mass star formation away from the well studied molecular clouds.  At a distance of 1000 pc, Class II YSOs less massive than $\sim0.5$ $M_\sun$ go undetected, but Class I and II detections are sufficient to give an overview of the broad picture of low-mass star formation across tens of square degrees.

\acknowledgments

This publication makes use of data products from the Wide-field Infrared Survey Explorer, which is a joint project of the University of California, Los Angeles, and the Jet Propulsion Laboratory/California Institute of Technology, funded by the National Aeronautics and Space Administration (NASA).  The work of WJF and MS was supported by appointments to the NASA Postdoctoral Program at Goddard Space Flight Center.  This research made use of Montage, which is funded by the National Science Foundation under Grant Number ACI-1440620, and was previously funded by NASA's Earth Science Technology Office, Computation Technologies Project, under Cooperative Agreement Number NCC5-626 between NASA and the California Institute of Technology.

\clearpage

\setcounter{figure}{6}
\begin{figure*}
\includegraphics[width=\hsize]{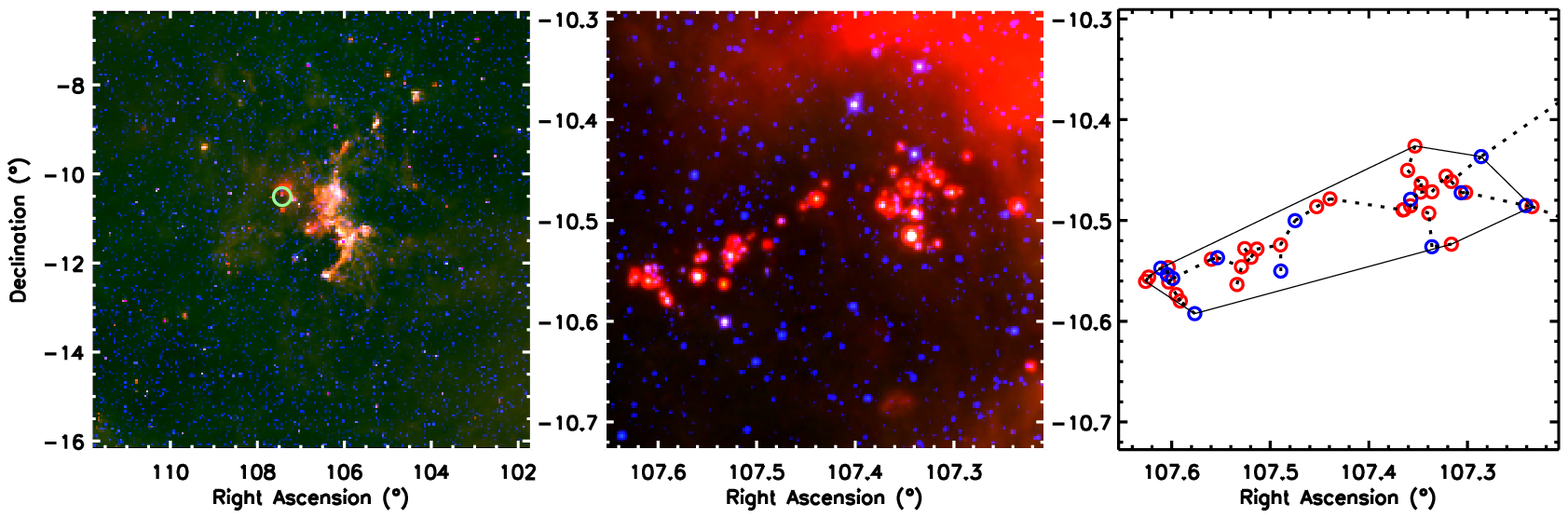}
\caption{Group 00.  {\em Left:} WISE map of the full 100 deg$^2$ region, where blue is 3.4 \micron, green is 12 \micron, red is 22 \micron, and the circle shows the location of the group.  {\em Center:} A closer look at the group, with the same assignment of color to wavelength.  At a distance of 1000 pc, the image is 7.6 pc wide. {\em Right:} A schematic view at the same scale as the previous panel. Red circles indicate Class I candidates; blue circles indicate Class II candidates.  The minimum spanning tree is shown with dotted lines, and the convex hull for the group is shown with a thin solid line.\label{f.cluster0}}
\end{figure*}

\begin{figure*}
\includegraphics[width=\hsize]{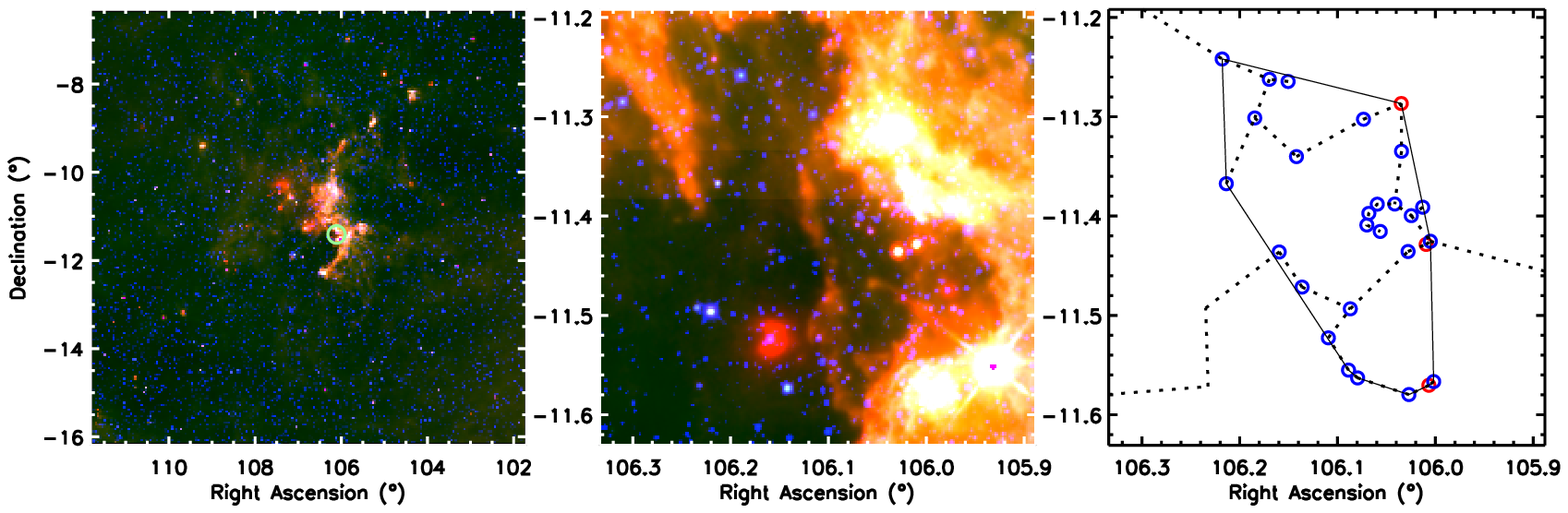}
\caption{Group 01.  See the caption to Figure~\ref{f.cluster0} for details.\label{f.cluster1}}
\end{figure*}

\begin{figure*}
\includegraphics[width=\hsize]{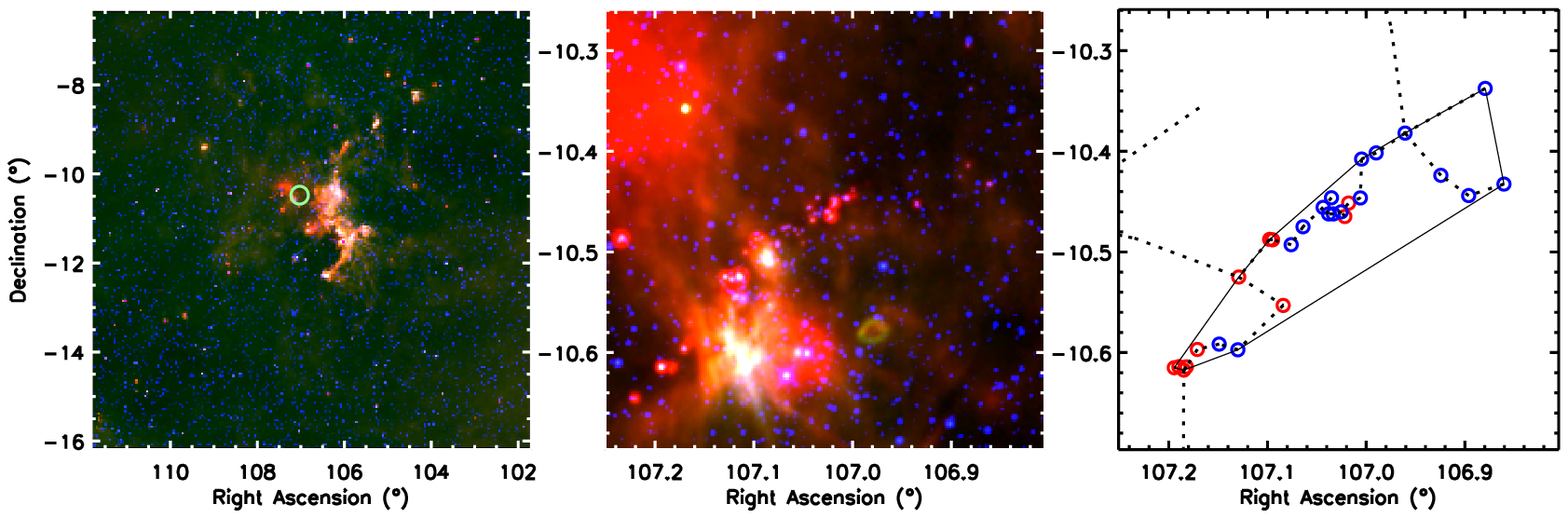}
\caption{Group 02.  See the caption to Figure~\ref{f.cluster0} for details.\label{f.cluster2}}
\end{figure*}

\begin{figure*}
\includegraphics[width=\hsize]{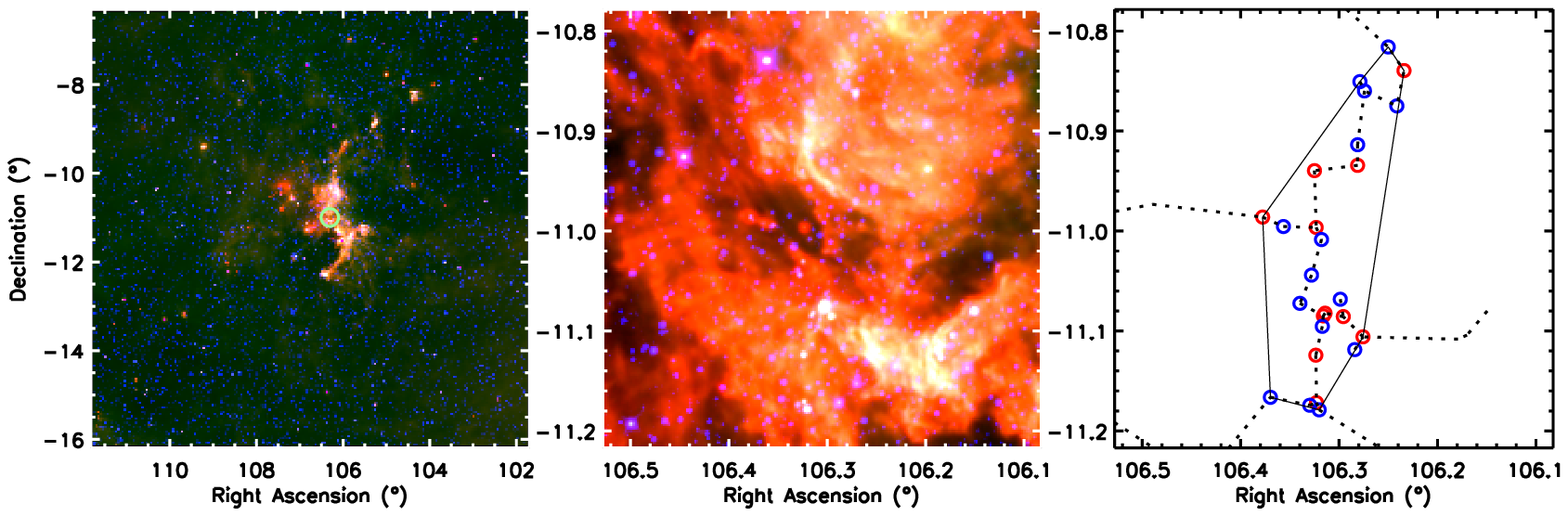}
\caption{Group 03.  See the caption to Figure~\ref{f.cluster0} for details.\label{f.cluster3}}
\end{figure*}

\begin{figure*}
\includegraphics[width=\hsize]{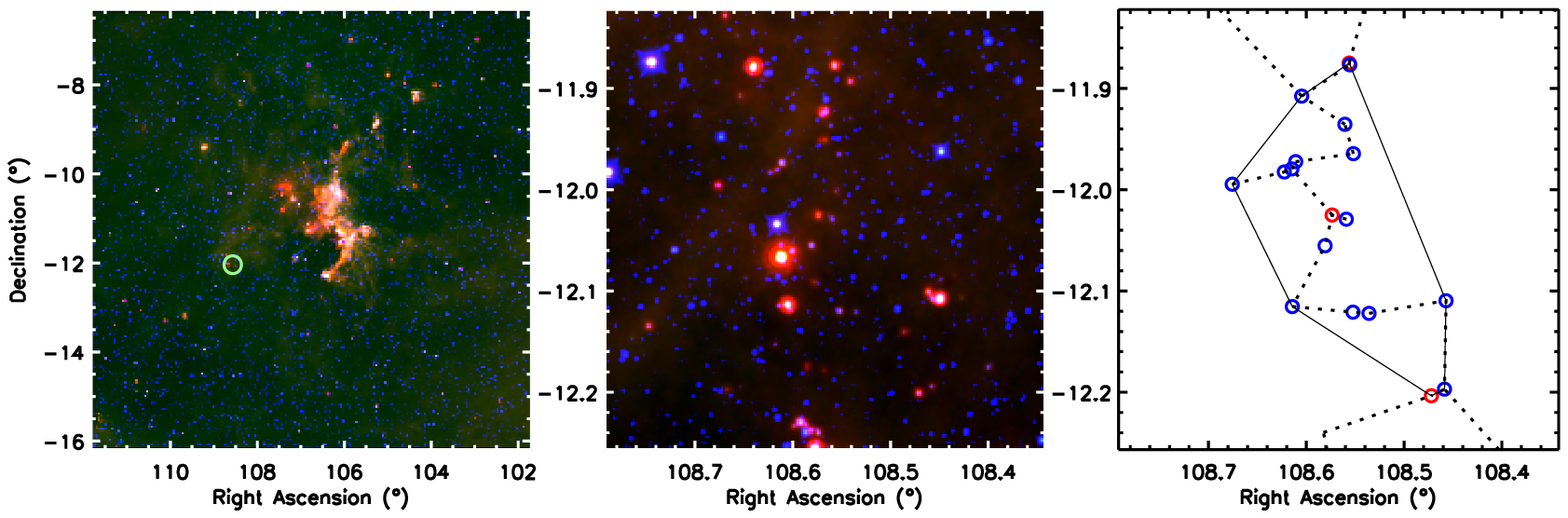}
\caption{Group 04.  See the caption to Figure~\ref{f.cluster0} for details.\label{f.cluster4}}
\end{figure*}

\begin{figure*}
\includegraphics[width=\hsize]{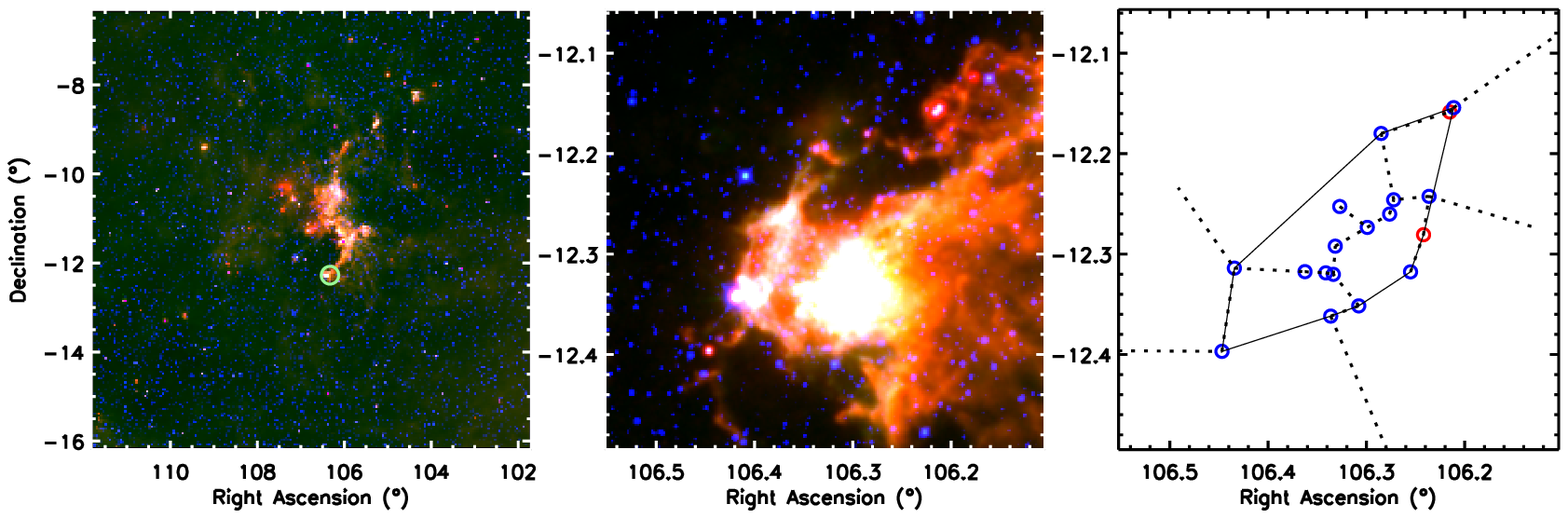}
\caption{Group 05.  See the caption to Figure~\ref{f.cluster0} for details.\label{f.cluster5}}
\end{figure*}

\begin{figure*}
\includegraphics[width=\hsize]{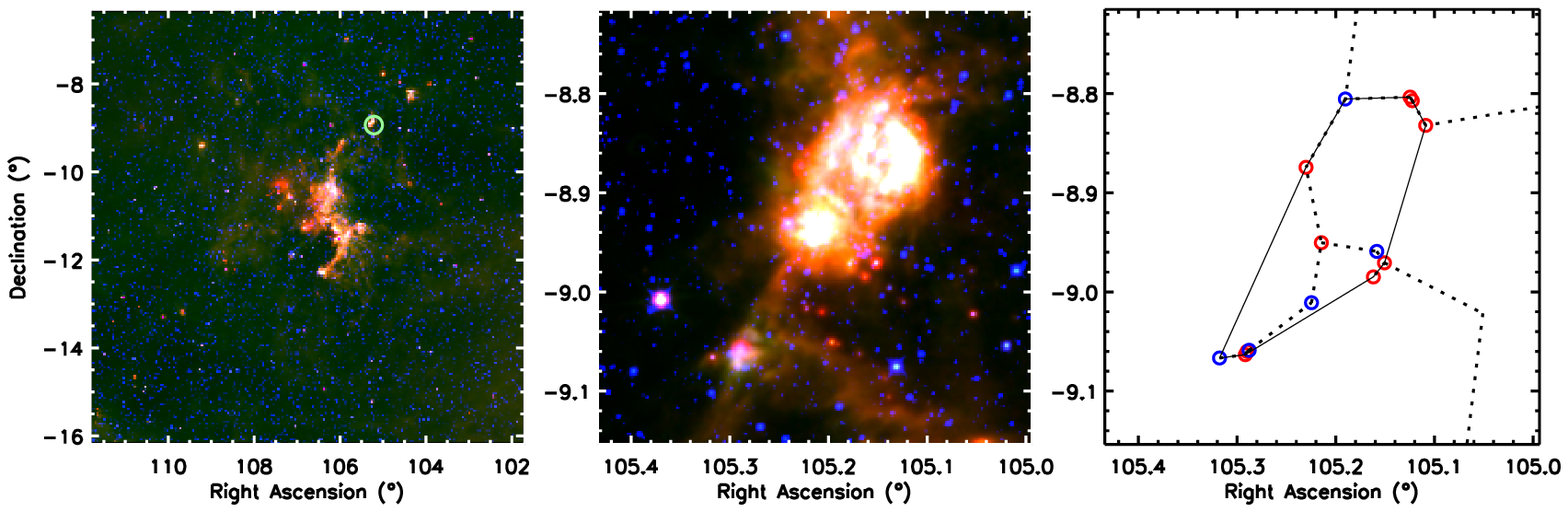}
\caption{Group 06.  See the caption to Figure~\ref{f.cluster0} for details.\label{f.cluster6}}
\end{figure*}

\begin{figure*}
\includegraphics[width=\hsize]{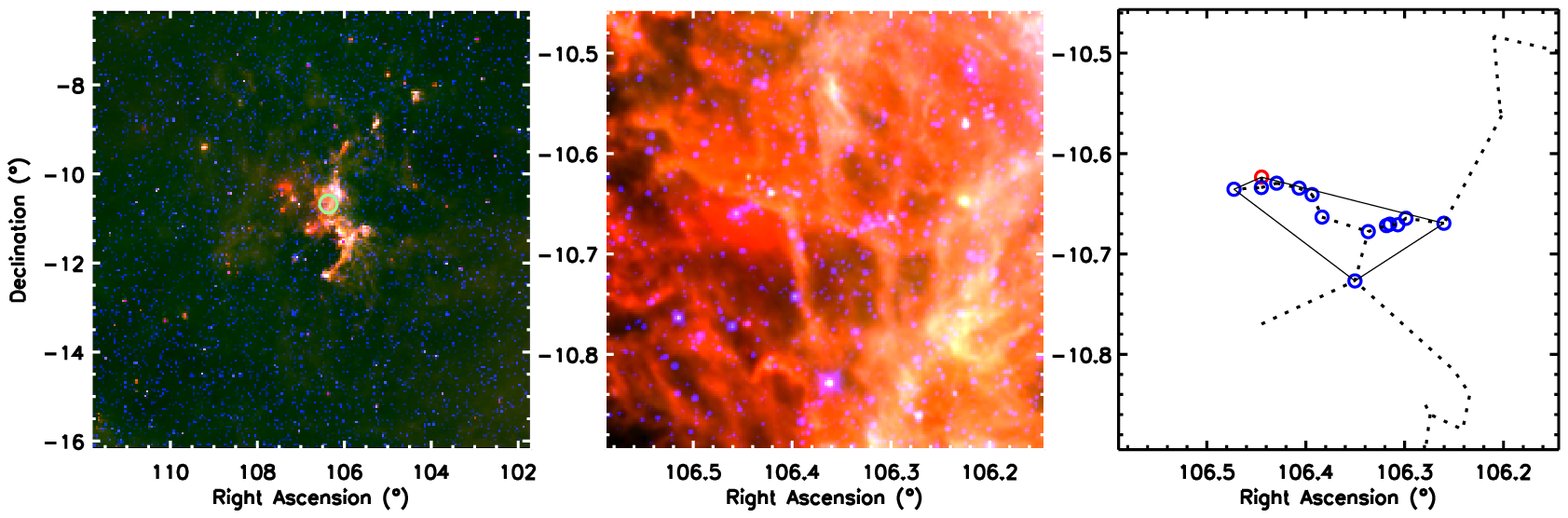}
\caption{Group 07.  See the caption to Figure~\ref{f.cluster0} for details.\label{f.cluster7}}
\end{figure*}

\begin{figure*}
\includegraphics[width=\hsize]{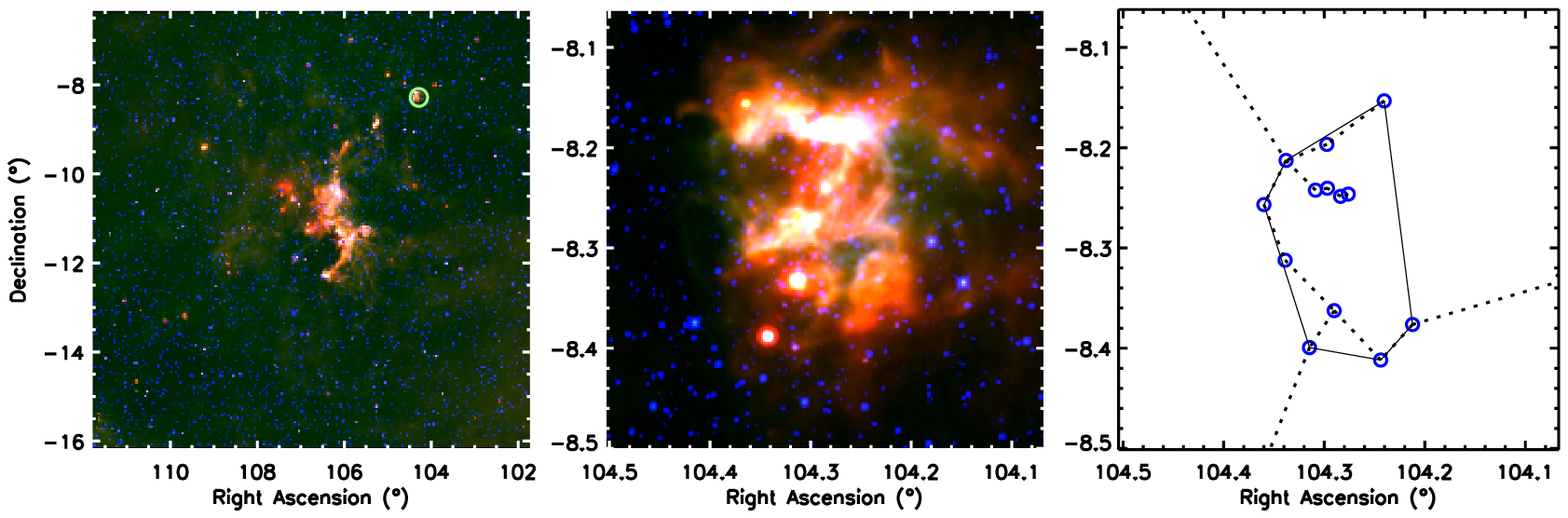}
\caption{Group 08.  See the caption to Figure~\ref{f.cluster0} for details.\label{f.cluster8}}
\end{figure*}

\begin{figure*}
\includegraphics[width=\hsize]{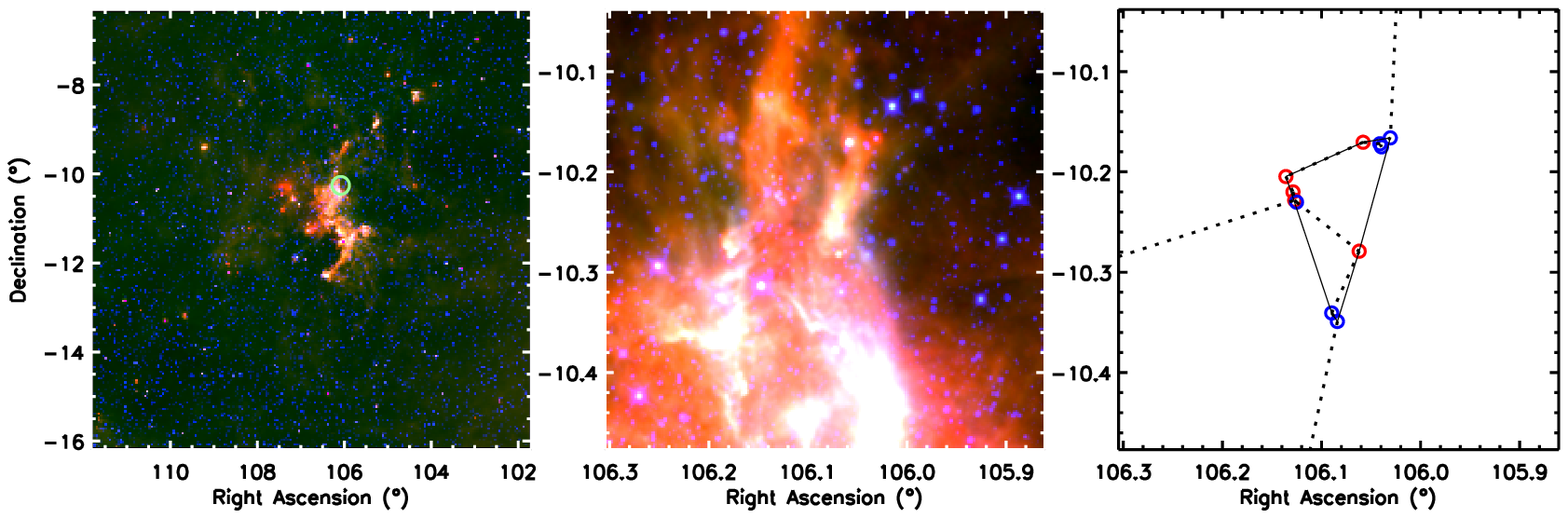}
\caption{Group 09.  See the caption to Figure~\ref{f.cluster0} for details.\label{f.cluster9}}
\end{figure*}

\begin{figure*}
\includegraphics[width=\hsize]{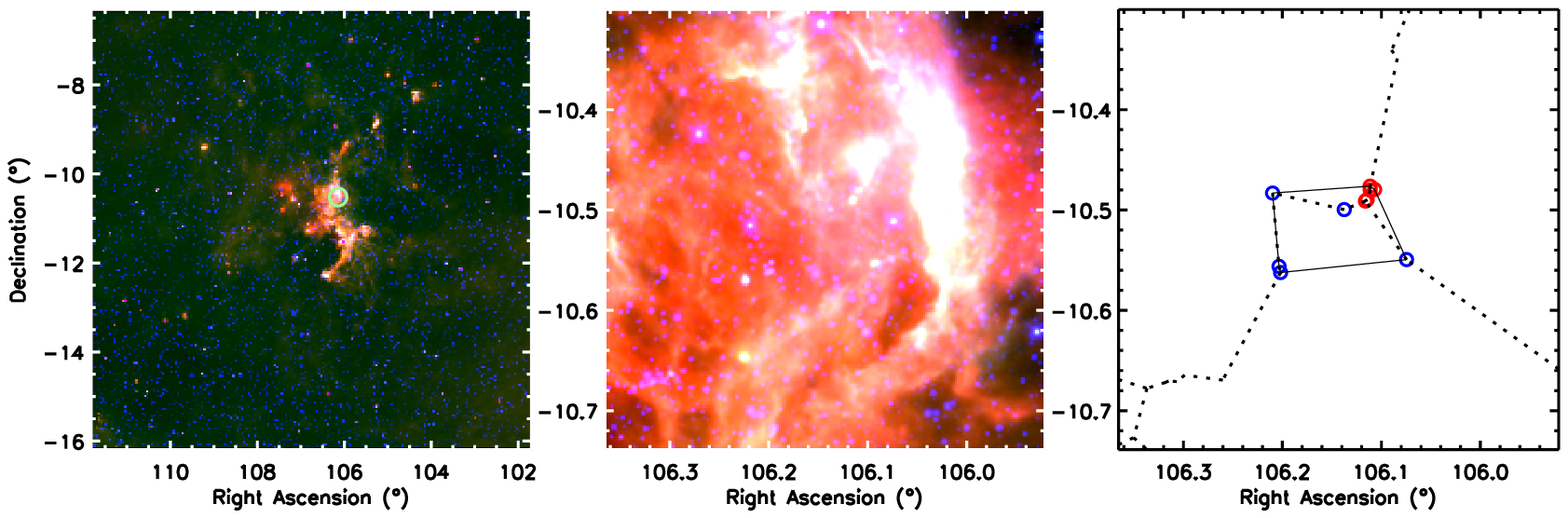}
\caption{Group 10.  See the caption to Figure~\ref{f.cluster0} for details.\label{f.cluster10}}
\end{figure*}

\begin{figure*}
\includegraphics[width=\hsize]{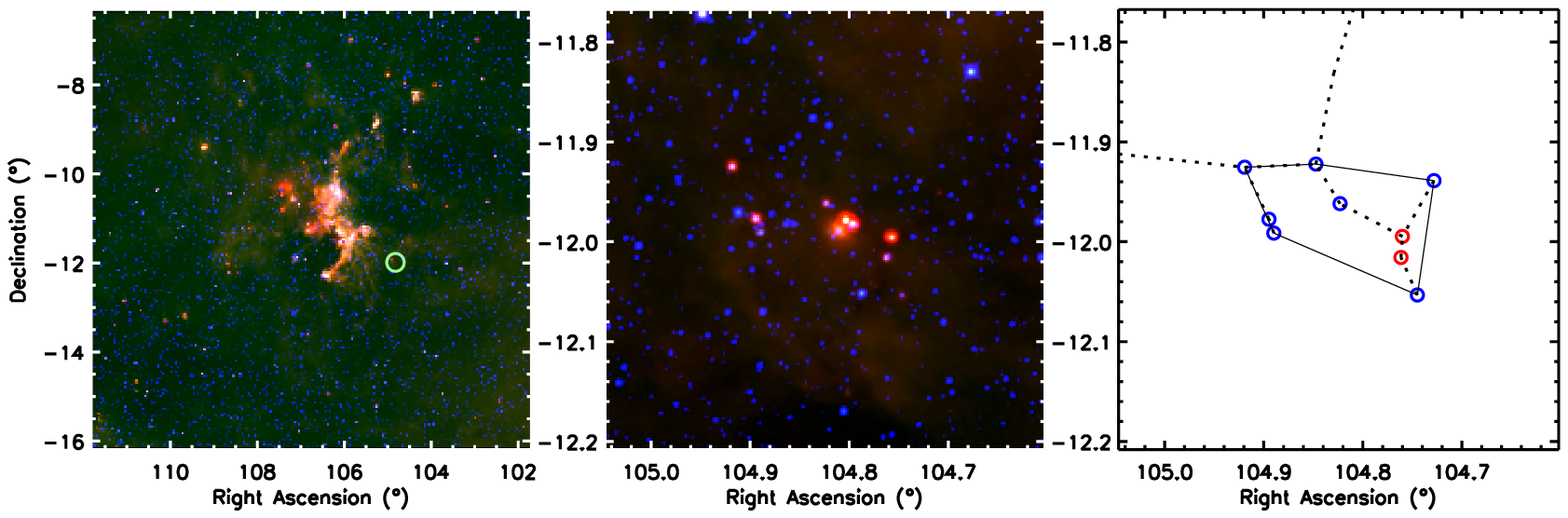}
\caption{Group 11.  See the caption to Figure~\ref{f.cluster0} for details.\label{f.cluster11}}
\end{figure*}

\begin{figure*}
\includegraphics[width=\hsize]{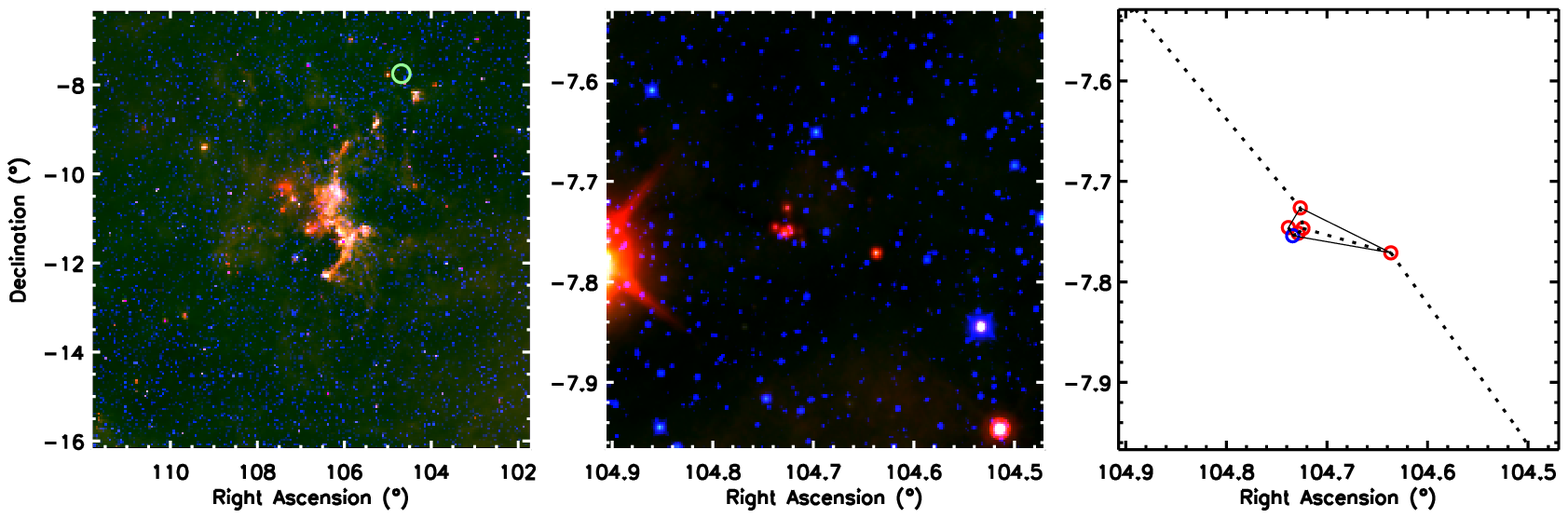}
\caption{Group 12.  See the caption to Figure~\ref{f.cluster0} for details.\label{f.cluster12}}
\end{figure*}

\begin{figure*}
\includegraphics[width=\hsize]{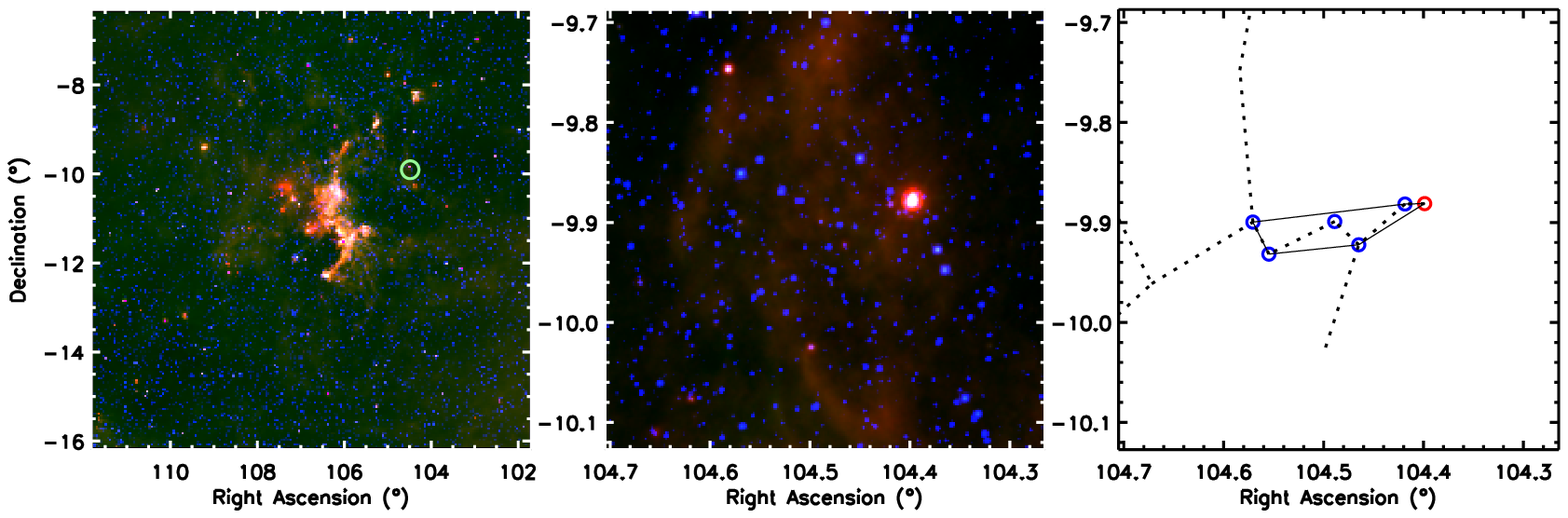}
\caption{Group 13.  See the caption to Figure~\ref{f.cluster0} for details.\label{f.cluster13}}
\end{figure*}

\begin{figure*}
\includegraphics[width=\hsize]{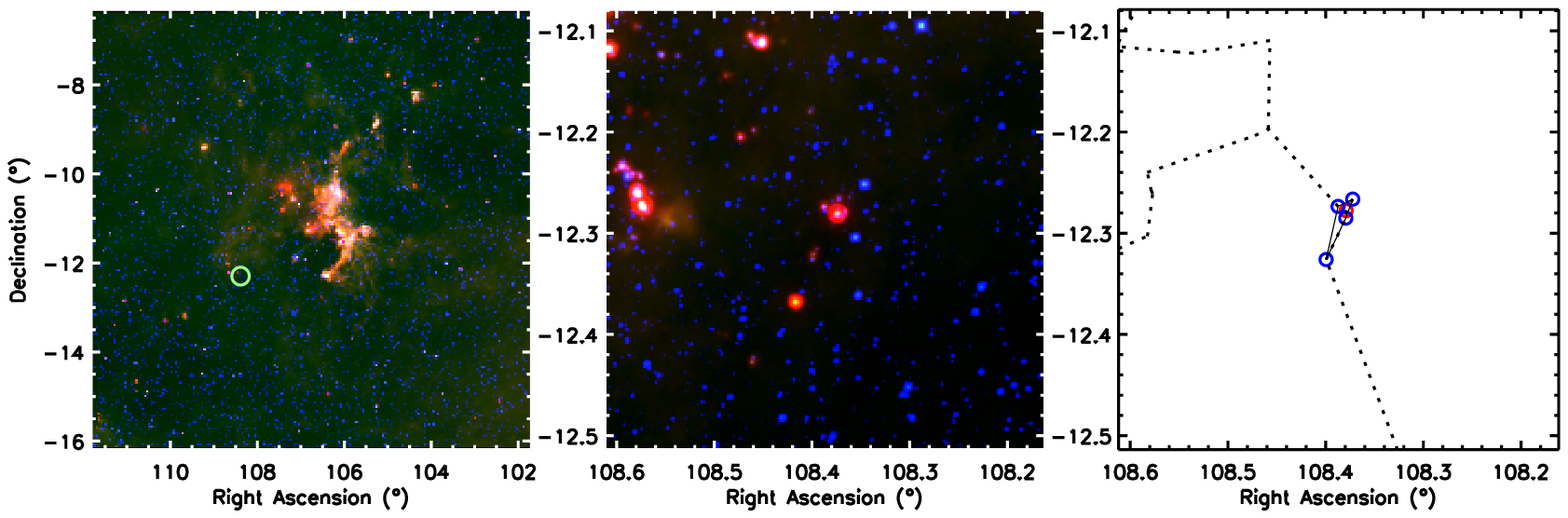}
\caption{Group 14.  See the caption to Figure~\ref{f.cluster0} for details.\label{f.cluster14}}
\end{figure*}

\begin{figure*}
\includegraphics[width=\hsize]{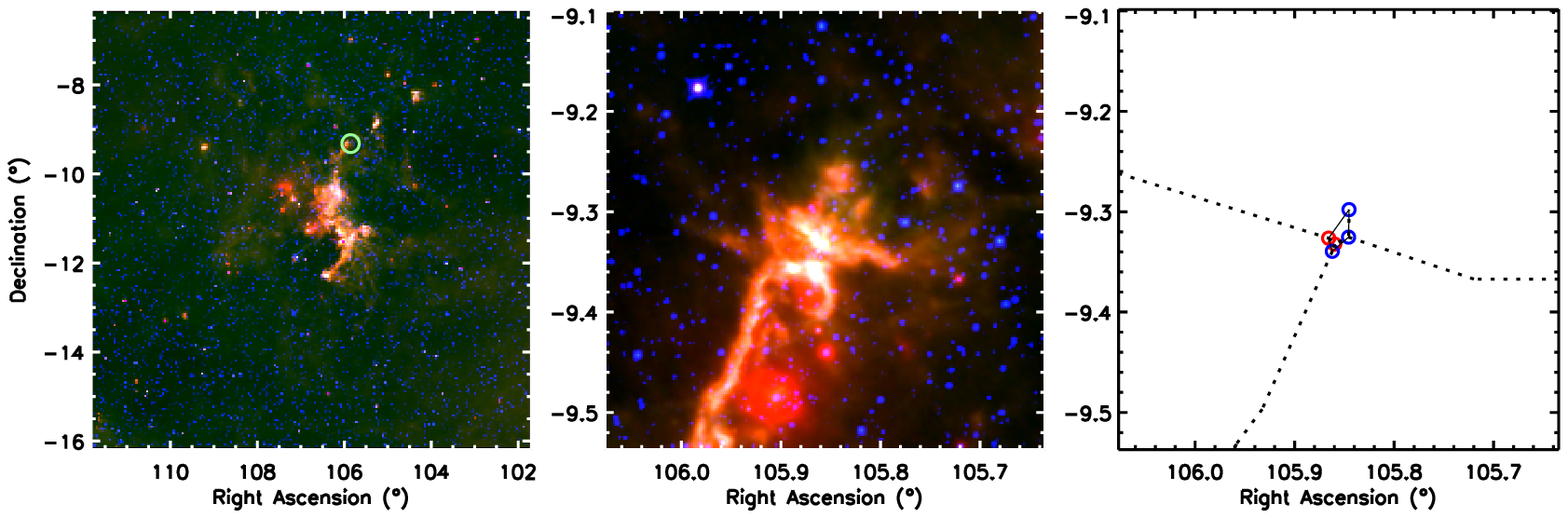}
\caption{Group 15.  See the caption to Figure~\ref{f.cluster0} for details.\label{f.cluster15}}
\end{figure*}

\renewcommand\thetable{1}
\input{yso_table.tex}

\end{document}

%% file: yso_table.tex
\clearpage
\LongTables
\tabletypesize{\tiny}
\begin{deluxetable}{lcrcccccccccccccccc}
\tablecaption{WISE YSO Candidates\tablenotemark{1}\label{t.yso}}
\tablehead{\colhead{WISE ID} & \multicolumn{2}{c}{Coordinates} & \colhead{$J$} & \colhead{$\sigma J$} & \colhead{$H$} & \colhead{$\sigma H$} & \colhead{$K$} & \colhead{$\sigma K$} & \colhead{$W1$} & \colhead{$\sigma W1$} & \colhead{$W2$} & \colhead{$\sigma W2$} & \colhead{$W3$} & \colhead{$\sigma W3$} & \colhead{$W4$} & \colhead{$\sigma W4$} & \colhead{Class} & \colhead{Code\tablenotemark{2}} \\
\cline{2-3} \\ [-1.5ex]
\colhead{} & \colhead{RA (deg)} & \colhead{Dec (deg)} & \colhead{(mag)} & \colhead{(mag)} & \colhead{(mag)} & \colhead{(mag)} & \colhead{(mag)} & \colhead{(mag)} & \colhead{(mag)} & \colhead{(mag)} & \colhead{(mag)} & \colhead{(mag)} & \colhead{(mag)} & \colhead{(mag)} & \colhead{(mag)} & \colhead{(mag)} & \colhead{} & \colhead{}}
\startdata
J064700.24$-$095944.9 & 101.75103 &  $-$9.99581 &  16.015 &   0.075 &  14.719 &   0.056 &  13.509 &   0.038 &  11.766 &   0.023 &  10.657 &   0.021 &   7.703 &   0.018 &   5.348 &   0.036 &  I & y \\
J064754.32$-$132823.5 & 101.97637 & $-$13.47320 &  14.560 &   0.035 &  13.826 &   0.036 &  12.813 &   0.027 &  11.259 &   0.022 &  10.317 &   0.018 &   7.381 &   0.016 &   5.609 &   0.035 & II & y \\
J064949.68$-$075032.8 & 102.45701 &  $-$7.84245 &  14.635 &   0.043 &  13.326 &   0.040 &  12.320 &   0.033 &  11.215 &   0.023 &  10.461 &   0.019 &   8.630 &   0.026 &   6.724 &   0.070 & II & y \\
J064958.55$-$073852.2 & 102.49397 &  $-$7.64784 &  11.233 &   0.027 &  10.315 &   0.026 &   9.416 &   0.023 &   8.119 &   0.022 &   7.376 &   0.019 &   4.329 &   0.015 &   2.244 &   0.020 & II & y \\
J065001.72$-$111814.1 & 102.50718 & $-$11.30393 &  12.594 &   0.024 &  11.531 &   0.024 &  10.283 &   0.019 &   8.208 &   0.023 &   6.844 &   0.023 &   4.015 &   0.015 &   2.643 &   0.019 &  I & y \\
J065007.04$-$074204.9 & 102.52934 &  $-$7.70137 &  11.757 &   0.024 &  10.813 &   0.026 &   9.999 &   0.021 &   9.084 &   0.023 &   8.564 &   0.020 &   6.397 &   0.016 &   4.835 &   0.028 & II & y \\
J065007.07$-$074010.4 & 102.52948 &  $-$7.66958 &  12.730 &   0.024 &  12.095 &   0.024 &  11.863 &   0.024 &  11.413 &   0.023 &  10.985 &   0.022 &   8.942 &   0.032 &   7.693 &   0.204 & II & y \\
J065010.78$-$073747.2 & 102.54492 &  $-$7.62980 &  12.141 &   0.024 &  11.948 &   0.026 &  11.726 &   0.024 &  11.107 &   0.023 &  10.514 &   0.021 &   8.199 &   0.024 &   4.998 &   0.034 & II & y \\
\enddata
\tablenotetext{1}{Cases with an entry for the magnitude and no entry for the uncertainty are upper limits.}
\tablenotetext{2}{Quality code from visual inspection:\ {\bf y}es; {\bf b}inary; {\bf f}aint; {\bf c}ontaminated; {\bf n}o. See Section \ref{s.filter} for details.}
\end{deluxetable}
\clearpage